# Annotation-Informed Block-Sparse Bayesian Modeling for cis-Expression Prediction

Lei Huang[1], Hui Shen[2], Kuan-Jui Su[2], Chuan Qiu[2], Martha Isabel Gonzalez-Ramirez[2], Anqi Liu[2], Zhe Luo[2], Yun Gong[2], Yipu Zhang[2], Dawei Li[3], Chaoyang Zhang[1,*], Hong-Wen Deng[2,*]

[1]School of Computing Sciences and Computer Engineering, University of Southern Mississippi, Hattiesburg, MS, USA, [2]Tulane Center for Biomedical Informatics and Genomics, Deming Department of Medicine, School of Medicine, Tulane University, New Orleans, LA, USA, [3]Texas Tech University Health Sciences Center, School of Medicine, Texas Tech University

*To whom correspondence should be addressed.

## Abstract

Genotype-based cis-expression prediction depends on accurately modeling local regulatory architecture. We present block-sparse Bayesian sparse linear mixed model (bsBSLMM), an extension of Bayesian sparse linear mixed model (BSLMM) that incorporates linkage disequilibrium (LD)-block spike-and-slab sparsity and a transcription start site (TSS)-informed SNP inclusion prior.

Across 23,098 genes from GEUVADIS European-ancestry lymphoblastoid cell lines, bsBSLMM retained more predictable genes than BSLMM, LASSO, BLUP, TIGAR elastic net, and TIGAR Dirichlet-process regression under matched evaluation criteria. Compared with BSLMM, bsBSLMM improved held-out prediction performance for most shared genes, with gains driven primarily by LD-block sparsity and further enhanced by the TSS-informed prior. Variants selected by bsBSLMM showed stronger enrichment in GM12878 DNase and H3K27ac regulatory regions than variants selected by BSLMM.

In transcriptome-wide association study (TWAS) analysis, bsBSLMM recovered established inflammatory bowel disease signals, including IL23R, and identified additional genome-wide significant genes not detected by BSLMM. Independent validation in the Louisiana Osteoporosis Study reproduced the increased prediction yield across ancestries and recovered biologically relevant bone mineral density pathways in downstream TWAS and gene set enrichment analyses.

These results demonstrate that incorporating LD-block structure and biologically informed SNP priors improves cis-expression prediction and enhances downstream TWAS discovery.

## 1. Introduction

Cis-genotype models are widely used to predict genetically regulated expression and to construct transcriptome-wide association study (TWAS) weights. PrediXcan [1], Bayesian sparse linear mixed model (BSLMM) [2], FUSION-style TWAS weights [3], best linear unbiased prediction (BLUP) [4], TIGAR elastic-net transcriptome imputation using the TIGAR framework [5] and elastic-net regularization [6], and TIGAR Dirichlet-process regression using latent Dirichlet-process regression [7] established the standard framework in which nearby variants are used to predict expression and fitted weights are subsequently evaluated by cross-validation, held-out prediction, or downstream gene-trait association. Summary-statistic TWAS using compatible prediction weights was later implemented in SPrediXcan [8]. Subsequent studies extended this framework through cross-tissue modeling, including UTMOST [9] and joint-tissue imputation [10], distal or multi-omic molecular information, including Bayesian genome-wide TWAS models using cis- and trans-eQTL information [11], MOSTWAS [12], and PUMICE [13], and deep sequence architectures, including Enformer [14] and Variformer [15]. These studies demonstrate that additional biological or statistical structure can improve prediction, although benchmark studies across diverse ancestries [16] indicate that no single approach is uniformly optimal across genes, tissues, ancestries, and regulatory architectures.

The underlying challenge is both statistical and biological. Reference transcriptome panels typically contain hundreds rather than tens of thousands of expression-profiled individuals, whereas each cis window may contain thousands of correlated variants. The resulting problem requires strong regularization of a high-dimensional genotype matrix while preserving sparse local effects that remain informative for gene-level prediction. At the same time, cis-regulatory architectures vary substantially across genes: some genes are driven by one dominant eQTL, others by multiple partially redundant effects in LD, and many by weak diffuse signal that is difficult to localize. Sparse regression, dense mixed models, and flexible Bayesian methods each capture part of this spectrum, but none explicitly models the local organization of correlated cis variants.

One underused source of structure in genotype-only expression prediction is local linkage disequilibrium (LD). SNPs within a cis window are ordered along the genome, clustered by local correlation, and frequently carry redundant regulatory information. Standard sparse priors largely treat SNPs as exchangeable after conditioning on the design matrix, whereas dense models distribute signal broadly across variants. Penalized regression methods shrink individual coefficients but do not explicitly encode a block-level null state. In practice, however, regulatory signal often arises within local haplotype neighborhoods, where excluding an entire LD segment may be more stable than independently suppressing many correlated SNP-level indicators.

Genomic position provides a second biologically relevant source of prior information. TSS-proximal variants are enriched for cis-regulatory activity in functional genomics resources such as ENCODE [17] and GTEx [18], although distal variants may also contribute through enhancers, chromatin contacts, or LD with causal variants. These observations motivate a prior that favors

TSS-proximal SNPs while allowing the data to determine the strength of this preference for each gene. In modestly sized reference panels, a learned TSS-distance prior is attractive because it avoids imposing a fixed external penalty that may not generalize across genes or tissues.

We propose block-sparse BSLMM (bsBSLMM), a hierarchical sparse mixed model designed around these cis-regulatory properties. bsBSLMM extends BSLMM with two components. First, contiguous LD blocks receive block-level indicators $z_b$, so that inactive blocks deterministically set all SNP effects within the block to zero. Second, the scalar SNP inclusion probability is replaced with a gene-specific TSS-distance prior, $\pi_j = \sigma(\alpha + \kappa a_j)$, learned jointly within the same MCMC chain. We evaluate bsBSLMM using GEUVADIS LCL expression from Battle et al. [19] and Lappalainen et al. [20], downstream IBD TWAS based on de Lange et al. [21], and independent Louisiana Osteoporosis Study (LOS) whole-blood validation [22] across two ancestries.

The study addresses three related questions. First, does an LD-block prior improve held-out cis-expression prediction relative to BSLMM and other widely used baselines? Second, do the block layer and TSS-distance prior provide separable contributions? Third, do the resulting posterior weights and active SNP sets recover independent regulatory annotations and disease-relevant TWAS signals? These questions are evaluated jointly because transcriptome-prediction weights are used not only for prediction but also for downstream biological interpretation.

## 2. Methods

### 2.1 Data and evaluation design

For each gene, biallelic SNPs within a $\pm 1$ Mb cis window were encoded as additive dosages and filtered for nonzero variance. Genotype columns and expression vectors were standardized within each cohort. GEUVADIS European-ancestry lymphoblastoid cell line (LCL) expression data from Battle et al. and Lappalainen et al. were used as the primary benchmark, with 23,098 autosomal genes included in the core bsBSLMM, BSLMM, LASSO [23], and BLUP analysis. Models were evaluated by 5-fold cross-validation on the development set; genes with CV-$r^2 \geq 0.01$ were subsequently refit on the full development set and assessed on held-out samples. The primary prediction filters required both CV-$r^2 \geq 0.01$ and held-out test $r > 0.4$, corresponding to an approximate two-sided Pearson-correlation threshold of $p = 0.011$ at the GEUVADIS held-out sample size. Full preprocessing details, gene counts, and per-chromosome diagnostics are provided in Supplementary Part II §S1–§S2.

The independent LOS validation used LOS whole-blood transcriptome and genotype data [22] from Caucasian American (CA) and African American (AA) sub-cohorts, aligned to GRCh38 [24]. All six methods were retrained within each ancestry using the same CV filter. To approximately match the GEUVADIS held-out significance level despite differing test sizes, LOS used cohort-specific held-out correlation filters: $r > 0.35$ in CA (median per-gene held-out $n = 53$) and $r >$ 0.43 in AA (median $n = 34$). LOS downstream transcriptome-wide association study (TWAS) analyses used LOS-trained PredictDB-format weights in SPrediXcan-style summary-statistic association against Morris et al. 2019 UK Biobank estimated bone mineral density (eBMD) genome-wide association study (GWAS) summary statistics [25].

Downstream biological analyses used independent regulatory annotations and GWAS summary statistics. Functional enrichment of credibly active SNPs was evaluated using ENCODE GM12878 DNase peaks and ENCODE GM12878 H3K27ac peaks as LCL-relevant chromatin annotations, together with GTEx v8 LCL cis-eQTL summaries as an external eQTL reference. GEUVADIS-trained TWAS weights were tested against de Lange et al. inflammatory bowel disease (IBD) GWAS summary statistics. LOS-trained TWAS weights were tested against Morris et al. UK Biobank eBMD GWAS summary statistics, followed by gene-set enrichment analysis (GSEA) [26] using curated bone-related gene sets.

Three complementary evaluation layers were used. First, retained-gene count measured how many genes passed both prediction filters, namely CV-$r^2 \geq 0.01$ and the cohort-specific held-out correlation threshold. Second, paired held-out accuracy compared methods on the same retained genes, using held-out Pearson $r$ and paired Wilcoxon tests across genes. These two analyses separate model coverage from per-gene predictive accuracy. Third, downstream biological utility was assessed through regulatory-annotation enrichment of credibly active SNPs and TWAS analyses of IBD and eBMD. These downstream analyses provide complementary evidence of biological plausibility and transferability.

**2.2 bsBSLMM model**

For standardized expression $\boldsymbol{y} \in \mathbb{R}^n$ and standardized cis-genotype matrix $\boldsymbol{X} \in \mathbb{R}^{n \times p}$, bsBSLMM uses

$$\boldsymbol{y} = \boldsymbol{X\beta} + \boldsymbol{g} + \boldsymbol{\varepsilon}, \qquad \boldsymbol{g} \sim \mathcal{N}(0, \sigma_\varepsilon^2 \eta_g \boldsymbol{K}), \qquad \boldsymbol{\varepsilon} \sim \mathcal{N}(0, \sigma_\varepsilon^2 \boldsymbol{I}),$$

where $\boldsymbol{K} = p^{-1} \boldsymbol{X} \boldsymbol{X}^\top$ is the cis-GRM. SNPs are partitioned into contiguous LD blocks by traversing position-ordered SNPs and starting a new block when the mean $r^2$ between the candidate SNP and the current block falls below 0.1, or when the block reaches 200 SNPs. Let $b(j)$ denote the block containing SNP $j$. bsBSLMM imposes

$$z_b \sim \text{Bernoulli}(\pi_{\text{block}}), \qquad \gamma_j \mid z_{b(j)} = 1 \sim \text{Bernoulli}(\pi_j), \qquad \pi_j = \sigma(\alpha + \kappa a_j),$$

with $\gamma_j = 0$ whenever $z_{b(j)} = 0$. Here $a_j = |\text{pos}_j - \text{TSS}|/10^6$ denotes SNP-to-TSS distance in megabases, with the TSS taken from the expression annotation. Conditional on inclusion, $\beta_j \sim \mathcal{N}(0, \sigma_\varepsilon^2 \eta_\beta)$; otherwise $\beta_j = 0$. The priors are $\pi_{\text{block}} \sim \text{Beta}(1,1)$, $\alpha \sim \mathcal{N}(\text{logit}(0.05), 4)$, $\kappa \sim \mathcal{N}(0,4)$, together with inverse-gamma priors on $\sigma_\varepsilon^2$, $\eta_\beta$, and $\eta_g$. Figure 1 summarizes the hierarchy, and full derivations and conditional updates are provided in Supplementary Methods §S1–§S3.

**Figure 1.** bsBSLMM model architecture. Contiguous cis-SNPs are partitioned into LD blocks. A block-level spike-and-slab indicator $z_b$ gates each block, and active blocks contain SNP-level indicators $\gamma_j$ with inclusion probabilities $\pi_j = \sigma(\alpha + \kappa a_j)$ learned from SNP-to-TSS distance. The model additionally includes a dense cis-GRM background effect.

## 2.3 Inference, prediction, and comparators

Posterior inference uses a Metropolis-within-Gibbs sampler. Each sweep samples block indicators, performs sequential SNP-level spike-and-slab scans within active blocks, samples the rotated dense random effect after eigendecomposition of the cis-GRM, updates variance scales, updates $(\alpha, \kappa)$ by random-walk Metropolis-Hastings, and samples $\pi_{\text{block}}$ from its beta full conditional. Real-data MCMC runs used 5,000 iterations, 2,500 burn-in iterations, and thinning interval 25. Held-out individual-level prediction used the sparse-cis predictor $\hat{y} = \boldsymbol{X}_{\text{test}}\hat{\beta}$ for bsBSLMM, BSLMM, and all baselines, matching the PredictDB weight format introduced by PrediXcan and used by SPrediXcan.

GEUVADIS comparisons included BSLMM, LASSO, BLUP, TIGAR elastic net (TIGAR EN) using the TIGAR framework and elastic-net regularization, TIGAR Dirichlet-process regression (TIGAR DPR) using latent Dirichlet-process regression in the TIGAR framework, block-only bsBSLMM, and full bsBSLMM. BSLMM and bsBSLMM used matched MCMC budgets. LASSO, BLUP, and TIGAR baselines followed their standard solver conventions under identical gene windows, sample splits, and held-out evaluation rules. Runtime was measured separately under single-threaded BLAS.

Posterior summaries were used for both prediction and interpretation. The posterior mean effect vector was used for held-out prediction and exported as the TWAS weight vector. Empirical 95% posterior intervals were used to define credibly active SNPs, with activity assigned only when the interval excluded zero. This active-set definition is stricter than a nonzero penalized-regression coefficient and links enrichment analysis to posterior uncertainty rather than solely to point estimates.

## 3. Results

### 3.1 Controlled simulations support the intended prior behavior

Semi-synthetic simulations reused real GEUVADIS chr20 genotype matrices and generated expression under sparse-high-SNR, sparse-low-SNR, and polygenic-dominant regimes. In the sparse-high-SNR setting, bsBSLMM achieved the highest mean 5-fold CV Pearson $r$ (0.605) and strongest coefficient recovery (correlation 0.353), ahead of BSLMM (0.598 and 0.295) and LASSO (0.542 and 0.209). In the sparse-low-SNR setting, predictive performance compressed across Bayesian mixed models, although coefficient recovery continued to favor bsBSLMM. In the polygenic-dominant setting, BLUP achieved the highest prediction accuracy, consistent with its dense ridge-like formulation. Full scenario tables are provided in Supplementary §S3.

### 3.2 bsBSLMM improves GEUVADIS prediction yield and paired held-out accuracy

Across all 22 autosomes, bsBSLMM retained the largest number of genes passing both prediction filters among six methods (3,518 genes; Table 1, Figure 2). This exceeded matched-MCMC BSLMM by 388 genes (3,518 vs 3,130; +12.4%) and also exceeded LASSO, BLUP, TIGAR EN, and TIGAR DPR under the same filtering criteria. On shared retained genes, bsBSLMM showed significantly higher paired held-out performance than every baseline by paired Wilcoxon testing. Relative to BSLMM, 1,998 of 2,928 shared genes favored bsBSLMM (68.2%), with mean $\Delta r = +0.0189$ and $p = 6.85 \times 10^{-113}$. Paired improvements were also significant relative to LASSO, BLUP, TIGAR EN, and TIGAR DPR.

Because BSLMM shares the same sparse mixed-model backbone, the matched-MCMC BSLMM comparison most directly isolates the effect of the proposed hierarchical prior. The remaining baselines represent complementary modeling regimes, including sparse penalized regression (LASSO), dense polygenic shrinkage (BLUP), and widely used transcriptome-imputation frameworks (TIGAR EN/DPR). bsBSLMM consistently showed higher paired held-out performance across all five baseline comparisons. TIGAR DPR retained the largest CV-only set but lost many genes after held-out filtering, whereas bsBSLMM achieved the largest validated retained-gene count.

**Table 1.** GEUVADIS prediction benchmark. The two prediction filters are CV-$r^2 \geq 0.01$ and held-out $r > 0.4$. Paired comparisons are bsBSLMM minus each baseline on the shared retained-gene set.

| Method | CV-only pass | Pass both filters | Shared with bsBSLMM | bsBSLMM wins | Mean $\Delta r$ | Wilcoxon $p$ |
|---|---|---|---|---|---|---|
| bsBSLMM | 12,236 | **3,518** | - | - | - | - |
| BSLMM | 11,750 | 3,130 | 2,928 | 1,998 (68.2%) | +0.0189 | $6.85 \times 10^{-113}$ |
| LASSO | 8,387 | 3,164 | 2,956 | 1,734 (58.7%) | +0.0083 | $4.6 \times 10^{-33}$ |
| BLUP | 9,364 | 2,194 | 2,004 | 1,672 (83.4%) | +0.0728 | $5.0 \times 10^{-223}$ |
| TIGAR EN | 11,121 | 3,129 | 2,809 | 1,650 (58.7%) | +0.0111 | $8.4 \times 10^{-30}$ |
| TIGAR DPR | 19,142 | 3,239 | 2,766 | 1,731 (62.6%) | +0.0208 | $1.7 \times 10^{-66}$ |

A
BSLMM
LASSO
BLUP
TIGAR EN
TIGAR DPR
bsBSLMM
Genes (cv-r² ≥ 0.01 ∩ test-r > 0.4)
Chromosome
B
0.568
0.575
0.545
0.572
0.557
0.571
Mean test r (cv-r² ≥ 0.01 ∩ test_r > 0.4)
BSLMM
LASSO
BLUP
TIGAR EN
TIGAR DPR
bsBSLMM

**Figure 2.** GEUVADIS genome-wide prediction benchmark under both prediction filters (CV $r^2 \geq 0.01$ and held-out $r > 0.4$), matching the convention used in Table 1. **Left**: per-chromosome counts of genes passing both filters for each of bsBSLMM, BSLMM, LASSO, BLUP, TIGAR EN, and TIGAR DPR. **Right**: per-method mean held-out test r computed over each method's own set of genes passing both filters.

### 3.3 Ablation separates block-level and TSS-prior contributions

The full benchmark establishes that bsBSLMM improves prediction performance, but does not independently isolate the contributions of each hierarchical component. We therefore evaluated a three-level ablation under the same 22-autosome prediction filters: matched BSLMM, block-only bsBSLMM, and full bsBSLMM. The BSLMM-to-block-only comparison isolates the LD-block gate, the block-only-to-full comparison isolates the learned TSS-distance prior, and the BSLMM-to-full comparison reflects their combined effect. For each contrast, Table 2 reports both paired prediction quality on shared retained genes and the net change in the number of genes passing both filters.

**Table 2.** GEUVADIS ablation under the two prediction filters.

| Comparison | Shared genes | Wins for newer model | Mean $\Delta r$ | Net retained-gene gain |
|---|---|---|---|---|
| block-only bsBSLMM vs BSLMM | 2,922 | 1,914 (65.5%) | +0.0155 | +265 |
| full bsBSLMM vs block-only | 3,248 | 1,861 (57.3%) | +0.0035 | +123 |
| full bsBSLMM vs BSLMM | 2,928 | 1,998 (68.2%) | +0.0189 | +388 |

Most of the observed improvement was attributable to the LD-block layer under the present ablation design. Relative to BSLMM, block-only bsBSLMM favored 65.5% of shared retained genes, increased mean held-out $r$ by +0.0155 and added 265 net retained genes. Adding the TSS-distance prior to the block-only model produced a smaller but consistent increment, yielding 57.3% wins, mean $\Delta r = +0.0035$, and 123 additional retained genes. The direct full-model comparison against BSLMM was approximately additive across these two components, with mean $\Delta r = +0.0189$ and 388 additional retained genes.

The pattern is consistent with distinct functional roles for the two hierarchy levels: the block gate changes the effective unit of sparsity from individual SNPs to LD neighborhoods, whereas the TSS prior redistributes SNP-level inclusion probability within active regions. The full ablation visualization is provided in Supplementary Figure S4.

### 3.4 Posterior summaries recover regulatory structure

Figure 3 summarizes the per-gene posterior decomposition (panels A–C) and the learned TSS-distance prior (panels D–F) across 23,098 GEUVADIS fits.

The architecture decomposition (Figure 3, A–C) shows three complementary views. Figure 3A plots the joint distribution of cis-$h^2$ (PVE) and the sparse share (PGE): 24.7% of genes have at least one credibly active SNP, with 18.6% sparse-dominant (PGE $\geq$ 0.5) and 75.3% polygenic-dominant (PGE < 0.1); among the active-SNP genes, the median sparse share is PGE = 0.62. Figure 3B compares posterior PVE to 5-fold CV $r^2$: most genes sit below the y = x identity, reflecting irreducible variance from train/test sampling, with the spread along the line tracking how much of the posterior heritability is recoverable as predictability. Figure 3C plots the posterior mean number of active blocks against CV r — the active-set size grows with predictability, consistent with the block layer concentrating signal in a small number of LD neighborhoods rather than across many isolated SNPs.

The TSS-prior validation (Figure 3, D–F) confirms that the learned annotation prior recovers a genuine TSS-proximal regulatory bias rather than imposing one a priori. Figure 3D shows the histogram of the learned TSS-distance coefficient $\hat{\beta}_{\text{coef}}$: it is negative in 22,812 of 23,098 genes (98.8%), with a median below zero, indicating that closer-to-TSS SNPs receive systematically higher prior inclusion probability when supported by the gene-specific data. Figure 3E plots $|\hat{\beta}|$ against $|\mathrm{SNP} - \mathrm{TSS}|$ for credibly active SNPs pooled from the top 500 strong-signal genes: effect-size mass concentrates near the TSS, with a long tail of distal active SNPs that the model retains when the genotype-expression likelihood supports them. Figure 3F connects the prior to prediction quality: per-gene $|\hat{\beta}_{\mathrm{coef}}|$ (TSS-prior strength) is positively associated with the bsBSLMM-vs-BSLMM held-out Δr, and the binned median Δr rises with prior strength, indicating that genes for which the TSS-distance signal is more decisive are the same genes for which bsBSLMM gains the most over BSLMM.

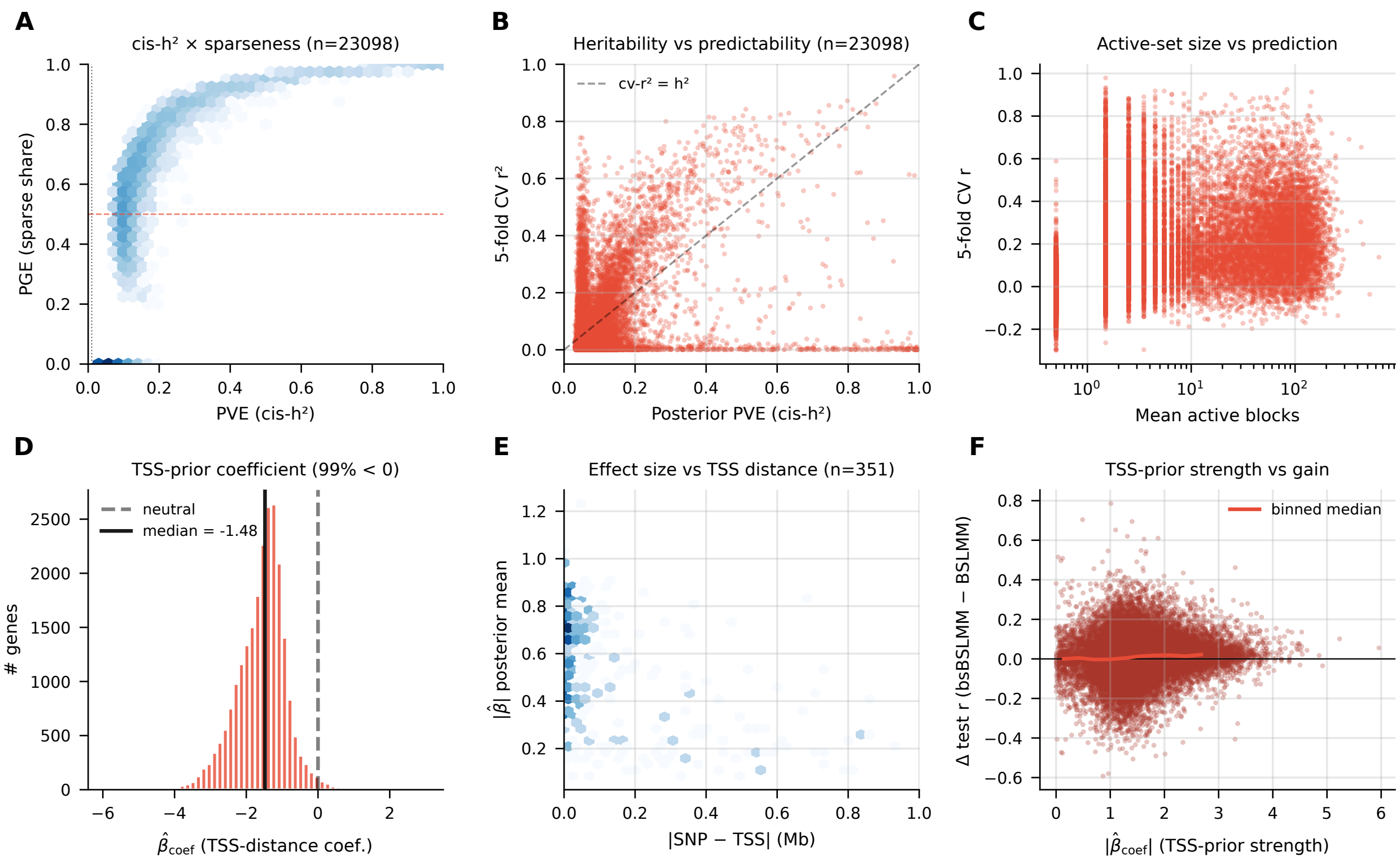

**Figure 3.** Posterior architecture and TSS-prior diagnostics across 23,098 GEUVADIS fits. **A**: joint distribution of cis-h² (PVE) and sparse share (PGE) — horizontal dashed line at PGE = 0.5 separates sparse-dominant from polygenic-dominant genes. **B**: posterior PVE vs 5-fold CV r²; the y = x line is the noise-free upper bound. **C**: posterior mean # active blocks vs CV r. **D**: histogram of the learned TSS-distance coefficient $\hat{\beta}_{\mathrm{coef}}$, with a median (solid) and the neutral $\beta_{\mathrm{coef}} = 0$ (dashed). **E**: pooled $|\hat{\beta}|$ vs $|\mathrm{SNP} - \mathrm{TSS}|$ for credibly active SNPs in the top-500 strong-signal genes; darker bins indicate higher local point density. **F**: per-gene $|\hat{\beta}_{\mathrm{coef}}|$ (TSS-prior strength) vs bsBSLMM-vs-BSLMM held-out Δr, with the binned median in red.

Credibly active SNPs also showed stronger functional enrichment than the BSLMM active set (Figure 4). bsBSLMM demonstrated the highest enrichment on the two LCL-relevant ENCODE chromatin tracks: 3.04-fold in GM12878 DNase peaks and 2.02-fold in H3K27ac peaks, compared with 1.40-fold and 1.35-fold for BSLMM. TIGAR EN showed the strongest enrichment in GTEx v8 LCL cis-eQTL replication, consistent with its similarity to GTEx elastic-net pipelines, whereas bsBSLMM showed the strongest enrichment on chromatin-mark annotations. The posterior-interval active set therefore prioritized variants overlapping independent regulatory evidence.

The posterior architecture also helps explain why predictive gains are not expected to be uniform across all genes. Many genes exhibit weak or diffuse cis signal at the present sample size, limiting the extent to which sparse priors can confidently localize effects. The observed improvements were concentrated in genes with detectable sparse or semi-sparse cis architecture, while the dense component absorbed background covariance. This interpretation is consistent with the benchmark results: bsBSLMM expanded validated coverage and improved paired prediction performance, although dense and penalized baselines remained competitive for some genes.

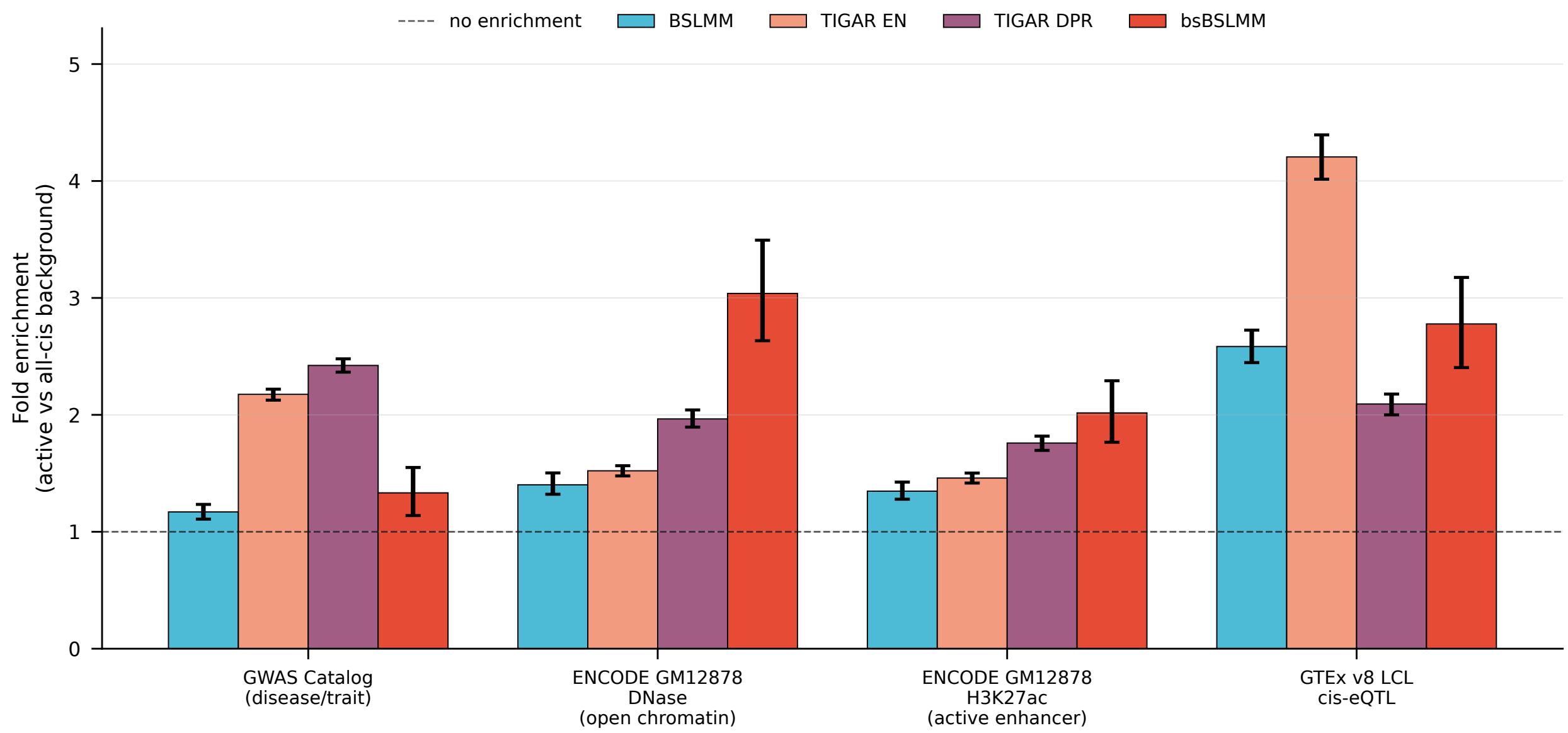

**Figure 4.** Functional enrichment of method-active cis-SNPs in external regulatory annotations. bsBSLMM active SNPs show the strongest enrichment in ENCODE GM12878 DNase and H3K27ac peaks, while TIGAR EN is strongest on GTEx v8 LCL cis-eQTL replication.

### 3.5 TWAS applications recover disease-relevant biology

Using GEUVADIS-trained weights in a SPrediXcan-style TWAS against de Lange et al. inflammatory bowel disease GWAS summary statistics, bsBSLMM tested 22,970 genes, recovered IL23R as the top signal ($p = 5.4 \times 10^{-47}$), and identified 30 genome-wide significant genes ($p < 5 \times 10^{-8}$) not reached by BSLMM on identical cis-SNP data at $\lambda_{\mathrm{GC}} = 1.36$ (Table 3). These included established immune or IBD-relevant loci highlighted by de Lange et al. and prior Crohn/IBD studies by Fransen et al. [27], Hampe et al. [28], Rioux et al. [29], and Liu et al. [30], such as PTPN2, ATG16L1, CISH, STAT5A, IL12RB2, UBA7, MSL1, LAMB1, CD19, NDFIP1, PARK7, and ZPBP2. Each unique bsBSLMM hit aggregated 1,717-5,101 allele-matched GWAS SNPs through the posterior-mean weight vector, and the signals spanned 12 chromosomes. TIGAR DPR produced more genome-wide significant hits overall, together with higher genomic inflation than bsBSLMM.

The IBD analysis evaluates the fitted weights in a downstream association setting. A method may improve expression prediction without producing stable gene-trait associations if its weights are poorly calibrated or overly diffuse. bsBSLMM increased tested-gene coverage relative to BSLMM, recovered canonical immune loci, and maintained lower genomic inflation than TIGAR EN and TIGAR DPR. The unique bsBSLMM signals were distributed across chromosomes and aggregated many allele-matched GWAS SNPs, supporting a gene-level imputed-expression interpretation.

### 3.6 LOS validation reproduces the prediction advantage across cohort and ancestry

The LOS whole-blood validation retrained all six methods in CA and AA cohorts. In CA, bsBSLMM retained the largest number of genes passing both filters (1,080 genes), exceeding BSLMM (955), TIGAR DPR (959), LASSO (903), TIGAR EN (852), and BLUP (548). Paired comparisons favored bsBSLMM over all five baselines at $p < 0.05$, including BSLMM ($\Delta = +0.0188, p = 1.2 \times 10^{-24}$), TIGAR DPR ($\Delta = +0.0115, p = 7.3 \times 10^{-9}$), and BLUP ($\Delta = +0.0442, p = 6.4 \times 10^{-25}$). In AA, bsBSLMM again retained the largest number of genes (722 vs 659 for BSLMM) and outperformed BSLMM, LASSO, and BLUP at $p < 0.05$; TIGAR comparisons remained positive but did not reach the same significance threshold.

LOS-trained TWAS against Morris eBMD GWAS summary statistics recovered curated BMD biology under both bsBSLMM and BSLMM (Table 3). After post-hoc genomic control, CA bsBSLMM produced 151 Bonferroni-significant genes versus 116 for CA BSLMM; AA bsBSLMM and AA BSLMM produced similar hit counts (177 and 181), with lower pre-calibration genomic inflation for AA bsBSLMM (1.90 vs 2.12). Preranked GSEA on signed genomic-controlled Z-scores recovered Morris-BMD and BMD-TWAS-colocalization gene sets [31], and the strongest population-matched signal was observed for AA bsBSLMM on the LOS structural-variant TWAS gene set (NES = -1.89, FDR $q = 0.001$). Additional LOS yield, H2H, TWAS, and GSEA tables are provided in the Supplementary Materials.

The LOS analysis changes cohort, tissue, sample size, and ancestry composition relative to GEUVADIS. The CA validation reproduced the GEUVADIS-scale bsBSLMM-versus-BSLMM paired effect in an independent whole-blood cohort, whereas the AA validation showed a smaller but still positive paired advantage relative to BSLMM and other non-TIGAR baselines. The eBMD TWAS and curated GSEA analyses further demonstrate that LOS-trained weights recover bone-relevant biology in a trait domain distinct from the GEUVADIS IBD application.

**Table 3.** Downstream biological validation summary.

| Analysis | bsBSLMM result | Comparator/context |
| --- | --- | --- |
| Functional enrichment using ENCODE and GTEx v8 | 3.04-fold GM12878 DNase; 2.02-fold H3K27ac | BSLMM: 1.40-fold DNase; 1.35-fold H3K27ac |

| Analysis | bsBSLMM result | Comparator/context |
|---|---|---|
| GEUVADIS-trained IBD TWAS using de Lange et al. GWAS | 22,970 genes tested; 76 genome-wide significant hits ($p < 5 \times 10^{-8}$); IL23R top hit ($p = 5.4 \times 10^{-47}$); 30 hits not reached by BSLMM | BSLMM: 21,920 genes, 72 hits; TIGAR DPR: 92 hits, higher $\lambda_{GC}$ (1.46 vs 1.36) |
| LOS CA eBMD TWAS using Morris et al. GWAS | 151 Bonferroni hits | CA BSLMM: 116 hits |
| LOS AA eBMD / GSEA using LOS and GSEA | 177 Bonferroni hits; G_LOS-SV-TWAS NES = -1.89, $q = 0.001$ | AA BSLMM: 181 hits; G_LOS-SV-TWAS did not pass the same FDR threshold |

### 3.7 Computational feasibility

The Python+Numba bsBSLMM sampler required a median full-fit time of 55.6 s/gene in the single-thread chr20 benchmark, compared with 34.4 s/gene for matched-MCMC BSLMM. A pybind11 C++ implementation of the per-iteration block-Gibbs sweep reduced this runtime to 30.8 s/gene while preserving the same statistical model and equivalent posterior summaries. LASSO and BLUP remained substantially faster, whereas bsBSLMM provided improved paired prediction performance, sparse posterior summaries, and TWAS-compatible weights. Full runtime tables are provided in the Supplementary Materials.

This runtime profile places bsBSLMM between fast point-estimation methods and more computationally intensive Bayesian workflows. LASSO and BLUP remain practical choices for large-scale screening analyses focused primarily on prediction. The native block-sweep implementation nevertheless makes bsBSLMM computationally comparable to matched-MCMC BSLMM while retaining the additional block and annotation hierarchy.

## 4. Discussion

bsBSLMM encodes two assumptions that are biologically plausible for cis-expression prediction: regulatory variants are organized by local LD structure, and TSS-proximal variants should receive higher prior inclusion probability when supported by the gene-specific data. The GEUVADIS benchmark indicates that this prior structure improves both validated coverage and paired held-out prediction relative to BSLMM and to sparse penalized, dense polygenic, and TIGAR-style baselines. Ablation analysis further suggests that the LD-block layer contributes most of the observed gain, whereas the TSS-distance prior provides an additional increase in retained-gene yield. This separation is statistically coherent because the block gate changes the effective unit of sparsity, whereas the TSS prior redistributes SNP-level inclusion probability within active regions.

The biological analyses support the posterior interpretation. bsBSLMM combines a smaller credibly active SNP set with stronger enrichment in LCL-relevant chromatin marks than the BSLMM active set. TWAS analyses further demonstrate that the resulting weights remain informative in downstream gene-trait association settings: the IBD analysis recovered canonical immune loci, whereas LOS-trained BMD analyses recovered curated bone biology with ancestry-specific patterns consistent with the validation cohort.

The study also highlights the importance of distinguishing yield from paired predictive performance. Marginal prediction summaries depend on the genes selected by each method and therefore conflate model quality with model selectivity. In contrast, paired comparisons isolate per-gene prediction differences on matched gene sets, while yield quantifies the number of genes retained after both CV and held-out filtering. bsBSLMM improved both quantities relative to BSLMM, supporting the interpretation that the block-sparse prior improves prediction rather than merely altering the evaluated gene set.

The active-SNP and TWAS analyses provide an additional layer of validation. A transcriptome-prediction method intended for TWAS should produce weights that are predictive, biologically plausible, and transferable to summary-statistic association analysis. bsBSLMM active SNPs showed enrichment in independent chromatin annotations, and its TWAS weights recovered known immune and bone biology across two training resources. Although these findings do not establish causal mechanisms for individual loci, they indicate that the posterior structure learned for prediction aligns with external regulatory and disease-relevance evidence.

Several limitations remain. The block on/off evidence uses a diagonal approximation within each block; although the subsequent within-block Gibbs scan respects LD through residual updates, the coarse block decision remains approximate. The production annotation is intentionally one-dimensional, relying only on TSS distance. Richer annotations, cross-gene sharing of $(\alpha, \kappa)$, cross-tissue priors, and ancestry-aware hierarchical extensions represent direct future extensions. Finally, the present study focuses on GEUVADIS LCL and LOS whole blood; tissue-matched training will likely be necessary for traits whose relevant cis regulation is concentrated in other tissues.

The current implementation additionally fits genes independently. This preserves compatibility with standard TWAS weight construction and enables direct comparison with existing per-gene methods, but it does not borrow information across genes with related architectures. Hierarchical extensions could share annotation effects, block-sparsity parameters, or variance components across genes, tissues, or ancestries. Such extensions may be particularly valuable in smaller reference panels, where gene-specific estimation of annotation effects becomes noisy. The logistic TSS prior also generalizes naturally to multiple annotations, including chromatin state, cCRE class, conservation, or tissue-matched accessibility.

In summary, bsBSLMM provides a modular Bayesian prior for genotype-based cis-expression prediction. It increased the number of GEUVADIS genes passing both prediction filters by 388 relative to matched BSLMM, showed higher paired held-out performance against all five baselines, produced biologically enriched sparse posterior effects, and reproduced the prediction advantage in an independent whole-blood cohort.

## 5. Data availability

This study used previously published or controlled-access data. Public inputs include GEUVADIS RNA-seq expression from Battle et al. and Lappalainen et al., 1000 Genomes Project Phase 3 genotypes [32], GENCODE v40 annotation [33], ENCODE GM12878 DNase tracks, ENCODE GM12878 H3K27ac tracks, GTEx v8 LCL cis-eQTL summaries, de Lange et al. IBD GWAS summary statistics, and Morris et al. eBMD GWAS summary statistics. LOS transcriptome and genotype data are controlled-access and available through the LOS investigators under data-use agreement. Derived tables and figure source data are reproducible from the repository paths mapped in the Supplementary Materials.

## 6. Code availability

Code is available at https://github.com/learnslowly/bsBSLMM under an MIT licence. The repository contains the bsBSLMM package, benchmark scripts, SLURM job wrappers, the native C++ block-sweep extension, and replication instructions. Source-data mappings are provided in the Supplementary Materials.

# Supplementary Materials for "Annotation-Informed Block-Sparse Bayesian Modeling for cis-Expression Prediction"

This supplement provides the full mathematical methods, secondary analyses, extended result tables, and source-data mappings supporting the main manuscript.

## Part I. Supplementary Methods

This section defines the full bsBSLMM probabilistic model, including the observed data, dense polygenic background, hierarchical sparse cis-effect prior, and hyperpriors. §S1 defines the model and posterior target; §S2 derives the conditional updates; and §S3 describes posterior summaries, prediction, and implementation.

The derivation follows the model's two genetic components. The vector $\boldsymbol{g}$ is a dense sample-level polygenic background governed by a Gaussian GRM prior, whereas $\boldsymbol{\beta}$ is a sparse SNP-level cis-effect vector governed by a block/SNP spike-and-slab prior. We first state the model, then derive the dense update for $\boldsymbol{g}$, the sparse updates for $(\boldsymbol{z}, \boldsymbol{\gamma}, \boldsymbol{\beta})$, and the hyperparameter updates.

### S1. Model definition and posterior target

This section provides the derivation-level counterpart of the manuscript Methods. It defines the observed gene-specific quantities, specifies the expression model as the sum of sparse cis-SNP effects, dense polygenic background, and residual noise, states the priors on the dense and sparse components, and then writes the posterior target used by the sampler.

For each target gene, let $\boldsymbol{X} \in \mathbb{R}^{n \times p}$ denote the standardized cis-genotype matrix for $n$ individuals and $p$ retained cis-SNPs, ordered by genomic position, and let $\boldsymbol{y} \in \mathbb{R}^{n}$ denote the standardized expression vector for the same individuals. Genotype columns and expression values were centered and scaled within each cohort. In the current implementation, the development cohort was standardized once before the 5-fold CV split, and the held-out test cohort was standardized independently using its own column statistics (scripts/run_bsbslmm_chunk.py and scripts/inference_bsbslmm.py). The CV stage is used as a screening step, whereas the held-out test evaluation is performed after refitting on the full development cohort. The exact CV and held-out prediction thresholds, their combined use, and test-cohort matching power are defined in Part II §S1.

LD blocks were constructed as described in the main Methods. SNPs were traversed in genomic order, and a new block was started whenever the mean squared correlation between candidate SNP $j$ and SNPs already in the current block fell below $r^2 < 0.1$, or when the current block reached 200 SNPs (bsbslmm/blocks.py::partition_blocks). After the initial greedy pass, any segment shorter than five SNPs was merged into the previous block to enforce a minimum block size of five SNPs, except for genes with fewer than five retained cis-SNPs after quality control.

The resulting LD partition is written as mutually exclusive and jointly exhaustive contiguous block index sets $\mathcal{J}_1, \dots, \mathcal{J}_B$:

$$\mathcal{J}_b \cap \mathcal{J}_c = \varnothing \ (b \neq c), \qquad \bigcup_{b=1}^{B} \mathcal{J}_b = \{1, \dots, p\}. \qquad (1)$$

For block $b$, let $k_b = |\mathcal{J}_b|$ and let $\boldsymbol{X}_b \in \mathbb{R}^{n \times k_b}$ denote the corresponding genotype submatrix. We write $b(j)$ for the unique block containing SNP $j$. Indices are used as follows: $j \in \{1, \dots, p\}$ indexes SNPs, $b \in \{1, \dots, B\}$ indexes LD blocks, $i \in \{1, \dots, n\}$ indexes individuals or rotated eigencomponents, and $m \in \{1, \dots, M\}$ indexes retained post-burn MCMC samples.

For each SNP $j$, the production annotation is the absolute distance to the gene's transcription start site (TSS) supplied by the expression annotation:

$$a_j = \frac{|\mathrm{pos}_j - \mathrm{TSS}|}{10^6}, \qquad (2)$$

measured in megabases. The genetic relationship matrix (GRM) computed from the standardized cis-genotype matrix is

$$\boldsymbol{K} = \frac{1}{p} \boldsymbol{X}\boldsymbol{X}^\top \in \mathbb{R}^{n \times n}. \qquad (3)$$

If $\boldsymbol{K}$ is rank-deficient, expressions involving $\boldsymbol{K}^{-1}$, $|\boldsymbol{K}|$, or eigenvalues are interpreted on the positive-eigenvalue subspace using the Moore–Penrose inverse and pseudo-determinant. In implementation, eigenvalues below $10^{-10}$ are treated as zero. After the eigendecomposition in §S2.2, define $n_{\mathrm{eff}} := |\{i : \lambda_i > 10^{-10}\}|$.

The variance notation is self-contained. The residual variance is $\sigma_\varepsilon^2$; the sparse-effect slab variance is $\sigma_\varepsilon^2 \eta_\beta$, where $\eta_\beta > 0$ is a variance ratio; and the dense polygenic background covariance is scaled by $\sigma_\varepsilon^2 \eta_g$, where $\eta_g > 0$ is another variance ratio. The symbols $\eta_\beta$ and $\eta_g$ are scale parameters, not indices.

### S1.1 Model definition

The model decomposes standardized expression into three components,

$$\boldsymbol{y} = \boldsymbol{X}\boldsymbol{\beta} + \boldsymbol{g} + \boldsymbol{\varepsilon}, \qquad (4)$$

where $\boldsymbol{\beta} = (\beta_1, \dots, \beta_p)^\top$ is the sparse SNP-level cis-effect vector, $\boldsymbol{g} \in \mathbb{R}^n$ is a dense sample-level polygenic background effect, and $\boldsymbol{\varepsilon} \in \mathbb{R}^n$ is residual noise. §S1.1.1–§S1.1.3 define the dense background prior, the hierarchical sparse prior, and the hyperpriors for the scale and inclusion parameters.

*S1.1.1 Observation model and dense polygenic background prior*

Conditional on the sparse cis effects and dense background, the observation model is Gaussian:

$$\boldsymbol{y} \mid \boldsymbol{\beta}, \boldsymbol{g}, \sigma_\varepsilon^2 \sim \mathcal{N}(\boldsymbol{X\beta} + \boldsymbol{g},\ \sigma_\varepsilon^2 \boldsymbol{I}_n), \qquad (5)$$

with density

$$p(\boldsymbol{y} \mid \boldsymbol{\beta}, \boldsymbol{g}, \sigma_\varepsilon^2) = (2\pi\sigma_\varepsilon^2)^{-n/2} \exp\left(-\frac{1}{2\sigma_\varepsilon^2} \|\boldsymbol{y} - \boldsymbol{X\beta} - \boldsymbol{g}\|_2^2\right). \qquad (6)$$

The dense background effect has a Gaussian GRM prior,

$$\boldsymbol{g} \mid \sigma_\varepsilon^2, \eta_g \sim \mathcal{N}\big(\boldsymbol{0},\ \sigma_\varepsilon^2 \eta_g \boldsymbol{K}\big), \qquad (7)$$

which gives the density

$$p(\boldsymbol{g} \mid \sigma_\varepsilon^2, \eta_g) = (2\pi\sigma_\varepsilon^2\eta_g)^{-n_{\mathrm{eff}}/2} |\boldsymbol{K}|_+^{-1/2} \exp\left(-\frac{1}{2\sigma_\varepsilon^2\eta_g}\ \boldsymbol{g}^\top \boldsymbol{K}^+ \boldsymbol{g}\right). \qquad (8)$$

Here $|\boldsymbol{K}|_+$ is the pseudo-determinant and $\boldsymbol{K}^+$ is the Moore–Penrose inverse. This dense component absorbs sample-level covariance induced by genetic relatedness or many weak background genetic contributions. It is distinct from the sparse cis-effect term $\boldsymbol{X\beta}$, which is designed to capture interpretable SNP-level effects in the local cis-window.

*S1.1.2 Hierarchical sparse cis-effect prior*

The sparse component is defined at the SNP level and organized by LD blocks. The prior contains two binary gates: a block-level indicator $z_b$ determines whether LD block $b$ may contain nonzero effects, and, conditional on an active block, a SNP-level indicator $\gamma_j$ determines whether SNP $j$ has a nonzero effect. The full generative specification is as follows.

At the block level,

$$z_b \mid \pi_{\text{block}} \sim \text{Bernoulli}(\pi_{\text{block}}), \qquad p(z_b \mid \pi_{\text{block}}) = \pi_{\text{block}}^{z_b}(1 - \pi_{\text{block}})^{1-z_b}, \quad b = 1, \dots, B. \qquad (9)$$

If $z_b = 0$, every SNP in block $b$ is deterministically inactive, so $\gamma_j = 0$ and $\beta_j = 0$ for all $j \in \mathcal{I}_b$.

At the SNP level,

$$\gamma_j \mid z_{b(j)}, \alpha, \kappa \sim \begin{cases} \text{Bernoulli}(\pi_j), & z_{b(j)} = 1, \\ 0, & z_{b(j)} = 0, \end{cases} \qquad \pi_j = \sigma(\alpha + \kappa\, a_j), \qquad (10)$$

where $\sigma(x) = 1/(1 + e^{-x})$ is the logistic link. The intercept $\alpha$ controls the baseline SNP-inclusion rate, and $\kappa$ is a global TSS-distance coefficient for the target gene. A negative $\kappa$ implies higher prior inclusion probability for variants closer to the TSS.

Given the SNP-level inclusion indicator, the sparse effect follows a spike-and-slab prior:

$$\beta_j \mid \gamma_j, \sigma_\varepsilon^2, \eta_\beta \sim \begin{cases} \mathcal{N}(0, \sigma_\varepsilon^2 \eta_\beta), & \gamma_j = 1, \\ \delta_0, & \gamma_j = 0, \end{cases} \qquad (11)$$

where $\delta_0$ denotes the degenerate point-mass distribution at zero. Equivalently,

$$p(\beta_j \mid \gamma_j, \sigma_\varepsilon^2, \eta_\beta) = \left[(2\pi\sigma_\varepsilon^2\eta_\beta)^{-1/2} \exp\left(-\frac{\beta_j^2}{2\sigma_\varepsilon^2\eta_\beta}\right)\right]^{\gamma_j} \left[\delta_0(\beta_j)\right]^{1-\gamma_j}. \qquad (12)$$

Marginalizing over the SNP-level indicator $\gamma_j$ while conditioning on an active parent block gives the spike-and-slab mixture prior $p(\beta_j \mid z_{b(j)} = 1, \ldots) = (1 - \pi_j)\,\delta_0(\beta_j) + \pi_j\,\mathcal{N}(\beta_j; 0, \sigma_\varepsilon^2\eta_\beta)$, used here only to interpret the sparse cis-effect. If $z_{b(j)} = 0$, then $p(\beta_j \mid z_{b(j)} = 0) = \delta_0(\beta_j)$. Thus the block gate supplies group sparsity across an LD region, while the SNP gate supplies annotation-informed sparsity within active regions.

*S1.1.3 Hyperpriors and default values*

The model is completed by assigning priors to the variance scales, the block-activation probability, and the annotation-prior coefficients. These priors regularize weakly identified genes and preserve closed-form Gibbs updates whenever conjugacy is available.

The variance-scale parameters have inverse-gamma priors,

$$\sigma_\varepsilon^2 \sim \text{InvGamma}(a_\varepsilon, b_\varepsilon), \qquad \eta_\beta \sim \text{InvGamma}(a_\beta, b_\beta), \qquad \eta_g \sim \text{InvGamma}(a_g, b_g). \qquad (13)$$

The defaults are

$$(a_\varepsilon, b_\varepsilon) = (1, 0.1), \qquad (a_\beta, b_\beta) = (a_g, b_g) = (2, 0.1). \qquad (14)$$

The residual-variance prior $\text{InvGamma}(1, 0.1)$ is broad on the standardized expression scale. It discourages pathological variance inflation while allowing the likelihood to determine $\sigma_\varepsilon^2$ for genes with appreciable signal. The sparse-effect and polygenic-background variance ratios, $\eta_\beta$ and $\eta_g$, share the default prior $\text{InvGamma}(2, 0.1)$, which is slightly more informative than the residual-variance prior and discourages runaway variance ratios when the data contain limited information about either component. Under the parameterization

$$\beta_j \mid \gamma_j = 1, \sigma_\varepsilon^2, \eta_\beta \sim \mathcal{N}(0, \sigma_\varepsilon^2\eta_\beta), \qquad \boldsymbol{g} \mid \sigma_\varepsilon^2, \eta_g \sim \mathcal{N}(\boldsymbol{0}, \sigma_\varepsilon^2\eta_g\boldsymbol{K}), \qquad (15)$$

this choice regularizes the sparse slab and dense background on the same residual-variance scale without favoring either component a priori. The posterior allocation between $\boldsymbol{X\beta}$ and $\boldsymbol{g}$ is therefore driven primarily by the likelihood.

The block-activation probability has a uniform Beta prior,

$$\pi_{\text{block}} \sim \text{Beta}(1,1). \qquad (16)$$

This prior is uniform on [0,1] and does not impose a preferred number of active LD blocks before observing the data. Conditional on the sampled block indicators, its posterior remains conjugate:

$$\pi_{\text{block}} \mid \boldsymbol{z} \sim \text{Beta}\left(1 + \sum_{b=1}^{B} z_b\,,\ 1 + B - \sum_{b=1}^{B} z_b\right). \qquad (17)$$

The annotation-prior coefficients have independent Gaussian priors,

$$\alpha \sim \mathcal{N}(\alpha_0, \tau_\alpha^2), \qquad \kappa \sim \mathcal{N}(0, \tau_\kappa^2), \qquad (18)$$

with default values $\alpha_0 = \text{logit}(0.05)$ and $\tau_\alpha^2 = \tau_\kappa^2 = 4$. The intercept prior centers the baseline SNP-level inclusion probability at 5% when the distance contribution is zero, since $\sigma(\text{logit}(0.05)) = 0.05$. This 5% value defines a weak sparse anchor on the logit scale. Effective sparsity is controlled jointly by the block gate $z_b$, the SNP-level gate $\gamma_j$, the sparse-effect variance ratio $\eta_\beta$, and the likelihood. Although the SNP-level prior begins in a low-inclusion regime, the posterior can move toward denser or sparser configurations when supported by the gene-specific data.

The prior variance $\tau_\alpha^2 = 4$ gives a prior standard deviation of 2 on the logit scale. The approximate 95% prior interval for $\alpha$ is therefore $\alpha_0 \pm 4$, which corresponds to a broad range of baseline inclusion probabilities. This prevents the intercept from being fixed near 5% while still regularizing genes with weak information about sparsity.

The prior for the global TSS-distance coefficient is centered at zero. Because

$$\pi_j = \sigma(\alpha + \kappa a_j), \qquad (19)$$

the value $\kappa = 0$ corresponds to no directional distance effect in the SNP-level inclusion prior. Negative posterior values of $\kappa$ imply that SNPs farther from the TSS have lower prior inclusion probability, whereas positive values imply that distal SNPs have higher prior inclusion probability. The default variance $\tau_\kappa^2 = 4$ allows substantial variation in the distance effect on the log-odds scale without forcing a TSS-proximal pattern.

The random-walk Metropolis-Hastings proposals for the annotation-prior coefficients use step sizes $(\delta_\alpha, \delta_\kappa) = (0.3, 0.5)$. These values were chosen to give stable acceptance rates in preliminary chromosome-level runs under the production MCMC schedule. They are proposal-scale parameters only; they do not define the prior and do not affect the posterior target.

### S1.2 Joint density and posterior target

Let $\boldsymbol{\theta} = (\sigma_\varepsilon^2, \eta_\beta, \eta_g, \pi_{\text{block}}, \alpha, \kappa)$ collect the hyperparameters. By the directed structure above, the joint density factors as

$$p(\boldsymbol{y}, \boldsymbol{\beta}, \boldsymbol{\gamma}, \boldsymbol{z}, \boldsymbol{g}, \boldsymbol{\theta}) = p(\boldsymbol{y} \mid \boldsymbol{\beta}, \boldsymbol{g}, \sigma_\varepsilon^2)\, p(\boldsymbol{g} \mid \sigma_\varepsilon^2, \eta_g)$$
$$\times \prod_{b=1}^{B} p\left(z_b \mid \pi_{\text{block}}\right) \prod_{j=1}^{p} p\left(\gamma_j \mid z_{b(j)}, \alpha, \kappa\right) \prod_{j=1}^{p} p\left(\beta_j \mid \gamma_j, \sigma_\varepsilon^2, \eta_\beta\right) p(\boldsymbol{\theta}). \tag{20}$$

The posterior target is therefore proportional to this joint density,

$$p(\boldsymbol{\beta}, \boldsymbol{\gamma}, \boldsymbol{z}, \boldsymbol{g}, \boldsymbol{\theta} \mid \boldsymbol{y}, \boldsymbol{X}, \{a_j\}_{j=1}^{p}, \{\mathcal{I}_b\}_{b=1}^{B}) \propto p(\boldsymbol{y}, \boldsymbol{\beta}, \boldsymbol{\gamma}, \boldsymbol{z}, \boldsymbol{g}, \boldsymbol{\theta}), \tag{21}$$

where the genotype matrix, TSS-distance annotations, and LD-block partition are treated as fixed inputs for each gene-specific fit.

For deriving Gibbs updates, first write $n_\gamma := \sum_{j=1}^{p} \gamma_j$ for the number of active SNPs and $\mathcal{A} := \{j : z_{b(j)} = 1\}$ for the active-block SNP set. The per-SNP indicator factor in eq. 20 decomposes as

$$\prod_{j=1}^{p} p\left(\gamma_j \mid z_{b(j)}, \alpha, \kappa\right) = \left(\prod_{j \in \mathcal{A}} \pi_j^{\gamma_j} \left(1 - \pi_j\right)^{1-\gamma_j}\right)\left(\prod_{j \notin \mathcal{A}} \mathbf{1}\left\{\gamma_j = 0\right\}\right), \tag{22}$$

where the second product enforces deterministic inactivity for SNPs in inactive blocks. The slab factor is

$$\prod_{j=1}^{p} p\left(\beta_j \mid \gamma_j, \sigma_\varepsilon^2, \eta_\beta\right) = \left(2\pi\sigma_\varepsilon^2\eta_\beta\right)^{-n_\gamma/2} \exp\left(-\frac{1}{2\sigma_\varepsilon^2\eta_\beta} \sum_{j:\gamma_j=1} \beta_j^2\right) \prod_{j:\gamma_j=0} \delta_0\left(\beta_j\right). \tag{23}$$

Taking logarithms and dropping additive constants independent of the unknowns gives

$$
\begin{aligned}
\log p(\boldsymbol{y}, \boldsymbol{\beta}, \boldsymbol{\gamma}, \boldsymbol{z}, \boldsymbol{g}, \boldsymbol{\theta}) \overset{c}{=} & -n/2 \log\sigma_\varepsilon^2 - 1/2\sigma_\varepsilon^2 \parallel \boldsymbol{y} - \boldsymbol{X}\boldsymbol{\beta} - \boldsymbol{g} \parallel_2^2 \\
& - n_{\text{eff}}/2 \log(\sigma_\varepsilon^2 \eta_g) - 1/2\sigma_\varepsilon^2 \eta_g \ \boldsymbol{g}^\top \boldsymbol{K}^+ \boldsymbol{g} \\
& + \left(\sum_{b=1}^{B} z_b\right) \log\pi_{\text{block}} + \left(B - \sum_{b=1}^{B} z_b\right) \log(1 - \pi_{\text{block}}) \\
& + \sum_{j \in \mathcal{A}} \left[\gamma_j \log\pi_j + (1 - \gamma_j) \log(1 - \pi_{\square\square})\right] \qquad (24) \\
& - n_\gamma/2 \log(\sigma_\varepsilon^2 \eta_\beta) - 1/2\sigma_\varepsilon^2 \eta_\beta \sum_{j:\gamma_j = 1} \beta_j^2 \\
& -(a_\varepsilon + 1)\log\sigma_\varepsilon^2 - b_\varepsilon/\sigma_\varepsilon^2 - (a_\beta + 1)\log\eta_\beta - b_\beta/\eta_\beta \\
& -(a_g + 1)\log\eta_g - b_g/\eta_g - (\alpha - \alpha_0)^2/2\tau_\alpha^2 - \kappa^2/2\tau_\kappa^2 \, .
\end{aligned}
$$

The symbol $\overset{c}{=}$ denotes equality up to an additive constant independent of the unknowns. eq. 24 is the source of the conditional updates derived in §S2: each update is obtained by retaining only the terms that involve the variable being updated.

## S2. Posterior inference: Metropolis-within-Gibbs sampler

### S2.1 Sampler overview and sweep order

Posterior inference is performed with a Metropolis-within-Gibbs sampler organized according to the model decomposition. The dense background $\boldsymbol{g}$ is updated through the GRM rotation derived in §S2.2. The sparse SNP effects $(\boldsymbol{\gamma}, \boldsymbol{\beta})$ are updated inside active LD blocks by the spike-and-slab Gibbs scan derived in §S2.3. The block indicators $\boldsymbol{z}$ are updated using the approximate collapsed block evidence in §S2.4. The variance-scale parameters, block-activation probability, and annotation-prior coefficients are updated in §S2.5.

Most updates are closed-form conditional draws. The annotation-prior coefficients $(\alpha, \kappa)$ require random-walk Metropolis-Hastings because $\pi_j = \sigma(\alpha + \kappa a_j)$ enters through a logistic link. The block on/off update uses a diagonal within-block approximation only for the coarse collapsed model-comparison step; after a block is active, the sequential within-block SNP scan updates the residual after each SNP draw and retains LD dependence through the current residual.

One sweep visits blocks first, then the dense background, then the variance-scale parameters, then the annotation-prior coefficients, and finally the block-activation probability. This ordering follows the implementation dependencies: the block-removed residual is reused inside the sparse scan, and the variance updates are delayed until the latent effects have been refreshed. §S2.2–§S2.5 derive these updates in sweep order, §S2.6–§S2.8 give the pseudocode, runtime equation manifest, and computational complexity, and §S3 covers posterior summaries, prediction, and the implementation footprint.

### S2.2 Dense polygenic update through GRM rotation

The polygenic vector $\boldsymbol{g}$ has a dense covariance matrix proportional to the GRM. Direct Gaussian sampling would require repeated dense linear algebra, whereas the rotation below diagonalizes this covariance once per gene. In the rotated basis, each component of $\tilde{\boldsymbol{g}}$ has an independent scalar Gaussian full conditional.

#### *S2.2.1 Rotation lemma*

The polygenic component $\boldsymbol{g}$ is the only continuous latent vector in the model whose prior eq. 7 has dense covariance; every other continuous latent variable ($\beta_j$ inside an active block, $\sigma_\varepsilon^2$, $\eta_\beta$, $\eta_g$) has scalar or diagonal structure. Sampling $\boldsymbol{g}$ directly would dominate per-sweep cost because each draw would solve an $n \times n$ linear system whose precision combines $\boldsymbol{K}^+/(\sigma_\varepsilon^2\eta_g)$ and $\boldsymbol{I}_n/\sigma_\varepsilon^2$, costing $O(n^3)$ per iteration. Across $T = 5{,}000$ sweeps and $n \approx 280$ samples, this would require $\sim 10^{11}$ operations per gene.

We use the standard mixed-model rotation: the GRM is diagonalized, and the polygenic update decouples into $n$ independent scalar Gaussian draws. The rotation matrix is the eigenbasis of $\boldsymbol{K}$ and is precomputed once per gene because $\boldsymbol{K}$ does not change across MCMC iterations. The rotated likelihood retains isotropic noise, so the per-SNP and block updates in §S2.3-S2.4 require no modification.

Diagonalize the GRM:

$$\boldsymbol{K} \;=\; \boldsymbol{U}\boldsymbol{\Lambda}\boldsymbol{U}^\top, \qquad \boldsymbol{U}^\top\boldsymbol{U} = \boldsymbol{I}_n, \qquad \boldsymbol{\Lambda} = \mathrm{diag}(\lambda_1, \dots, \lambda_n). \tag{25}$$

Define rotated quantities $\tilde{\boldsymbol{y}} = \boldsymbol{U}^\top\boldsymbol{y}$, $\tilde{\boldsymbol{X}} = \boldsymbol{U}^\top\boldsymbol{X}$, $\tilde{\boldsymbol{g}} = \boldsymbol{U}^\top\boldsymbol{g}$. Because $\tilde{\boldsymbol{g}}$ is an orthogonal-linear function of a Gaussian, it is itself Gaussian with mean $\boldsymbol{0}$ and covariance

$$\mathrm{Cov}(\tilde{\boldsymbol{g}}) \;=\; \boldsymbol{U}^\top(\sigma_\varepsilon^2\eta_g\boldsymbol{K})\boldsymbol{U} \;=\; \sigma_\varepsilon^2\eta_g\,\boldsymbol{U}^\top\boldsymbol{U}\boldsymbol{\Lambda}\boldsymbol{U}^\top\boldsymbol{U} \;=\; \sigma_\varepsilon^2\eta_g\boldsymbol{\Lambda}. \tag{26}$$

In components, $\tilde{g}_i \sim \mathcal{N}(0, \sigma_\varepsilon^2\eta_g\lambda_i)$ independently across $i$. The rotated likelihood retains the same isotropic-noise form, since $\boldsymbol{U}^\top(\sigma_\varepsilon^2\boldsymbol{I}_n)\boldsymbol{U} = \sigma_\varepsilon^2\boldsymbol{I}_n$ and

$$\tilde{\boldsymbol{y}} - \tilde{\boldsymbol{X}}\boldsymbol{\beta} - \tilde{\boldsymbol{g}} \;\sim\; \mathcal{N}(\boldsymbol{0}, \sigma_\varepsilon^2\boldsymbol{I}_n). \tag{27}$$

#### *S2.2.2 Component-wise polygenic posterior*

Conditional on the cis effects $\boldsymbol{\beta}$ and the noise variance $\sigma_\varepsilon^2$, define the rotated cis-residual $r_i := \tilde{y}_i - \tilde{\boldsymbol{X}}_{i,:}\boldsymbol{\beta}$. The conditional log-density of $\tilde{g}_i$ is, from eq. 26 and eq. 27,

$$\log p(\tilde{g}_i \mid \cdot) \;=\; -1/2\sigma_\varepsilon^2\,(r_i - \tilde{g}_i)^2 - 1/2\sigma_\varepsilon^2\eta_g\lambda_i\;\tilde{g}_i^2 + \mathrm{const}. \tag{28}$$

Expanding and collecting quadratic and linear terms in $\tilde{g}_i$,

$$\log p(\tilde{g}_i \mid \cdot) \; = \; -1/2 \; \tilde{g}_i^2\left(1/\sigma_\varepsilon^2 + 1/\sigma_\varepsilon^2\eta_g\lambda_i\right) + \tilde{g}_i \; r_i/\sigma_\varepsilon^2 + \text{const.} \qquad (29)$$

This is the canonical log density of a univariate Gaussian. Reading off the posterior variance and mean,

$$v_{g,i} \; = \frac{\sigma_\varepsilon^2\eta_g\lambda_i}{1+\eta_g\lambda_i}, \qquad \mu_{g,i} \; = v_{g,i} \cdot \frac{r_i}{\sigma_\varepsilon^2} \; = \frac{\eta_g\lambda_i}{1+\eta_g\lambda_i}\, r_i, \qquad (30)$$

hence

$$\tilde{g}_i \mid \cdot \sim \; \mathcal{N}\left(\eta_g\lambda_i/1+\eta_g\lambda_i \; r_i, \; \sigma_\varepsilon^2\eta_g\lambda_i/1+\eta_g\lambda_i\right). \qquad (31)$$

Each component $\tilde{g}_i$ is conditionally independent of the others given $\boldsymbol{\beta}$, so the full $\tilde{\boldsymbol{g}}$ update consists of $n$ independent scalar Gaussian draws: $O(n)$ per sweep rather than $O(n^3)$ for repeated dense inversion in the original basis. The factor $\eta_g\lambda_i/(1+\eta_g\lambda_i)$ multiplying $r_i$ is the BLUP shrinkage coefficient for eigencomponent $i$, equal to the prior-to-total variance ratio in that direction. Eigencomponents with $\eta_g\lambda_i \gg 1$ are weakly shrunk, whereas components with $\eta_g\lambda_i \ll 1$ are strongly shrunk.

The eigendecomposition is computed once per gene at startup (bsbslmm/bslmm_mcmc.py::_eigendecompose_K), and subsequent draws are produced by bsbslmm/bslmm_mcmc.py::_sample_g_rot. All subsequent updates operate in the rotated basis using the running residual $\tilde{\boldsymbol{r}} := \tilde{\boldsymbol{y}} - \tilde{\boldsymbol{X}}\boldsymbol{\beta} - \tilde{\boldsymbol{g}}$. The per-SNP and block on/off updates require inner products of the form $\tilde{\boldsymbol{X}}_{:,j}^\top \boldsymbol{r}_{-j}$ and squared norms $\| \boldsymbol{r}^{(-b)} \|_2^2$, which are invariant under the orthogonal rotation $\boldsymbol{U}^\top$ after precomputation.

**S2.3 Sparse cis-effect updates inside active LD blocks**

The per-SNP update determines which SNPs inside an active block carry nonzero effects and dominates per-sweep wall-clock time for genes with at least one active block. Because the slab prior on $\beta_j$ is conjugate to the Gaussian likelihood and $\gamma_j$ is binary, the joint conditional posterior on $(\gamma_j, \beta_j)$ has closed form: a Bernoulli mass on $\gamma_j$ whose log-odds equal the prior log-odds plus a Bayes factor, and, when $\gamma_j = 1$, a Gaussian density on $\beta_j$ with the standard single-SNP ridge-regression mean and variance.

The update is performed as a sequential scan. SNPs in an active block are visited in genomic order, and after each $(\gamma_j, \beta_j)$ draw the running residual $\tilde{\boldsymbol{r}}$ is updated before the next SNP is processed. Thus each SNP is conditioned on the freshly sampled effects of its block-mates through the residual, retaining within-block LD dependence during the scan.

For each SNP $j$ inside an active block ($z_{b(j)} = 1$), we update $(\gamma_j, \beta_j)$ jointly. Define the per-SNP residual that adds SNP $j$'s current contribution back into $\tilde{\boldsymbol{r}}$,

$$\boldsymbol{r}_{-j} \;:=\; \tilde{\boldsymbol{r}} + \widetilde{\boldsymbol{X}}_{:,j}\, \beta_j^{\mathrm{old}}. \qquad (32)$$

The joint conditional posterior on $(\gamma_j, \beta_j)$ factorizes into a Bernoulli mass on $\gamma_j$ and a Gaussian density on $\beta_j$ given $\gamma_j = 1$.

*S2.3.1 Conditional posterior of $\beta_j$ given inclusion*

When $\gamma_j = 1$, the rotated likelihood contribution to $\beta_j$, treating $\boldsymbol{r}_{-j}$ as the pseudo-observation, is proportional to

$$\exp(-\,1/2\sigma_\varepsilon^2 \parallel \boldsymbol{r}_{-j} - \widetilde{\boldsymbol{X}}_{:,j}\beta_j \parallel_2^2). \qquad (33)$$

Expanding the squared norm,

$$\parallel \boldsymbol{r}_{-j} - \widetilde{\boldsymbol{X}}_{:,j}\beta_j \parallel_2^2 \;=\; \parallel \boldsymbol{r}_{-j} \parallel_2^2 - 2\beta_j\, \widetilde{\boldsymbol{X}}_{:,j}^\top \boldsymbol{r}_{-j} + \beta_j^2\, \widetilde{\boldsymbol{X}}_{:,j}^\top \widetilde{\boldsymbol{X}}_{:,j}. \qquad (34)$$

Combining with the slab prior $\beta_j \mid \gamma_j = 1 \sim \mathcal{N}(0, \sigma_\varepsilon^2 \eta_\beta)$ and dropping the $\beta_j$-independent term,

$$\begin{aligned} &\log p(\beta_j \mid \gamma_j = 1, \boldsymbol{r}_{-j}, \cdot) \\ &= \;-\,1/2\;\beta_j^2\big(\widetilde{\boldsymbol{X}}_{:,j}^\top \widetilde{\boldsymbol{X}}_{:,j}/\sigma_\varepsilon^2 + 1/\sigma_\varepsilon^2\eta_\beta\big) + \beta_j\; \widetilde{\boldsymbol{X}}_{:,j}^\top \boldsymbol{r}_{-j}/\sigma_\varepsilon^2 + \mathrm{const.} \qquad (35) \end{aligned}$$

This is the canonical log density of a univariate Gaussian. Reading off the precision and mean,

$$v_j^{-1} \;=\; \frac{\widetilde{\boldsymbol{X}}_{:,j}^\top \widetilde{\boldsymbol{X}}_{:,j}}{\sigma_\varepsilon^2} + \frac{1}{\sigma_\varepsilon^2 \eta_\beta}, \qquad \mu_j \;=\; v_j \cdot \frac{\widetilde{\boldsymbol{X}}_{:,j}^\top \boldsymbol{r}_{-j}}{\sigma_\varepsilon^2}, \qquad (36)$$

hence

$$\beta_j \mid \gamma_j = 1, \boldsymbol{r}_{-j}, \cdot \;\sim\; \mathcal{N}(\mu_j,\, v_j). \qquad (37)$$

*S2.3.2 Bayes factor for inclusion*

To sample $\gamma_j$ we need the marginal likelihood ratio between $\gamma_j = 1$ and $\gamma_j = 0$:

$$\mathrm{BF}_j \;=\; \frac{p(\boldsymbol{r}_{-j} \mid \gamma_j = 1)}{p(\boldsymbol{r}_{-j} \mid \gamma_j = 0)} \;=\; \frac{\int p(\boldsymbol{r}_{-j} \mid \beta_j)\, \mathcal{N}(\beta_j; 0, \sigma_\varepsilon^2\eta_\beta)\, \mathrm{d}\beta_j}{p(\boldsymbol{r}_{-j} \mid \beta_j = 0)}. \qquad (38)$$

The numerator combines the likelihood eq. 6 with the slab density. After cancelling the $\beta_j$-independent factor $(2\pi\sigma_\varepsilon^2)^{-n/2}\exp(-\parallel \boldsymbol{r}_{-j} \parallel_2^2/(2\sigma_\varepsilon^2))$, the integrand is a Gaussian in $\beta_j$ whose log is exactly eq. 35. Completing the square in eq. 35,

$$-\,1/2v_j\, \beta_j^2 + \mu_j/v_j\, \beta_j \;=\; -\,1/2v_j\, (\beta_j - \mu_j)^2 + \mu_j^2/2v_j\,, \qquad (39)$$

and integrating the Gaussian kernel gives $\int \exp(-(\beta_j - \mu_j)^2/(2v_j))\, \mathrm{d}\beta_j = \sqrt{2\pi v_j}$. Collecting the surviving normalizers,

$$\int p(\boldsymbol{r}_{-j} \mid \beta_j)\, \mathcal{N}(\beta_j; 0, \sigma_\varepsilon^2 \eta_\beta)\, \mathrm{d}\beta_j \;=\; \frac{e^{-\|\boldsymbol{r}_{-j}\|_2^2/(2\sigma_\varepsilon^2)}}{(2\pi\sigma_\varepsilon^2)^{n/2}} \cdot \frac{\sqrt{2\pi v_j}}{\sqrt{2\pi\sigma_\varepsilon^2 \eta_\beta}} \cdot e^{\mu_j^2/(2v_j)}. \qquad (40)$$

The denominator of eq. 38 is the likelihood at $\beta_j = 0$,

$$p(\boldsymbol{r}_{-j} \mid \beta_j = 0) \;=\; \frac{1}{(2\pi\sigma_\varepsilon^2)^{n/2}}\, e^{-\|\boldsymbol{r}_{-j}\|_2^2/(2\sigma_\varepsilon^2)}. \qquad (41)$$

Dividing eq. 40 by eq. 41, the residual factors cancel exactly and the Bayes factor reduces to the closed form

$$\mathrm{BF}_j \;=\; \sqrt{\frac{v_j}{\sigma_\varepsilon^2 \eta_\beta}}\, e^{\mu_j^2/(2v_j)}, \qquad \mathrm{logBF}_j \;=\; 1/2\, \log v_j - 1/2\, \log(\sigma_\varepsilon^2 \eta_\beta) + \mu_j^2/2v_j\,. \qquad (42)$$

The first two terms constitute the marginal-likelihood penalty for the larger prior support of the slab. The third term, $\mu_j^2/(2v_j)$, rewards evidence for a nonzero effect through the squared posterior mean divided by twice the posterior variance. Together these terms yield the standard Bayesian model-selection trade-off between effect-size evidence and model complexity.

The Bayes factor is recomputed for every SNP at every Gibbs sweep. For a typical gene with $p \approx 3000$ cis-SNPs and $T = 5{,}000$ sweeps, eq. 42 is evaluated about $1.5 \times 10^7$ times per gene. The closed form is therefore essential for computational feasibility.

*S2.3.3 Posterior inclusion probability and joint draw*

Combining the per-SNP prior $\Pr(\gamma_j = 1 \mid z_{b(j)} = 1) = \pi_j$ with the Bayes factor eq. 42,

$$\Pr(\gamma_j = 1 \mid \boldsymbol{r}_{-j}, \cdot) \;=\; \frac{\pi_j\, \mathrm{BF}_j}{\pi_j\, \mathrm{BF}_j + (1 - \pi_j)}. \qquad (43)$$

For numerical stability we work in log-space: let $L_1 := \log\pi_j + \mathrm{logBF}_j$ and $L_0 := \log(1 - \pi_j)$, then $\Pr(\gamma_j = 1 \mid \cdot) = \sigma(L_1 - L_0) = 1/(1 + e^{L_0 - L_1})$.

Sample $\gamma_j$ from the resulting Bernoulli, and conditionally on $\gamma_j$ sample $\beta_j$ from eq. 37 when $\gamma_j = 1$ or fix $\beta_j = 0$ when $\gamma_j = 0$. Update the running residual in place,

$$\tilde{\boldsymbol{r}} \;\leftarrow\; \boldsymbol{r}_{-j} - \widetilde{\boldsymbol{X}}_{:,j}\, \beta_j^{\mathrm{new}}, \qquad (44)$$

and proceed to the next SNP. Because the scan is sequential, later SNPs in the block condition on the freshly updated effects of earlier SNPs via the running residual. Implementation: bsbslmm/_mcmc_kernel.py::gibbs_scan_block_inplace_e13.

### S2.4 Approximate collapsed LD-block update

The SNP-level scan in §S2.3 determines which SNPs carry nonzero effects conditional on an active block. The block-level update instead determines whether an entire LD region should be active. Directly sampling this coarse indicator conditional on the current SNP effects would mix poorly, so the update compares the block-off and block-on states after integrating over block-local SNP effects.

The block on/off update samples each block-level indicator $z_b$ conditional on the rest of the chain, with the per-SNP indicators $\boldsymbol{\gamma}_b$ and slab effects $\boldsymbol{\beta}_b$ for that block integrated out. This collapsed update permits transitions between inactive and active block states even when all SNP effects in an inactive block are deterministically zero.

The exact block-on marginal likelihood in eq. 47 sums over $2^{k_b}$ active-set configurations and is intractable for nontrivial block sizes. We therefore use a diagonal approximation only for this marginal model-comparison step: $\widetilde{\boldsymbol{X}}_b^\top \widetilde{\boldsymbol{X}}_b$ is replaced by its diagonal, reducing the sum to $k_b$ independent two-term contributions. Once a block is active, the within-block per-SNP scan in §S2.3 is run without this diagonal approximation and retains LD coupling through the sequential residual update.

#### *S2.4.1 Marginal likelihood under each block setting*

To sample $z_b$ we need the marginal likelihood of the residual under each setting, having integrated out $(\boldsymbol{\beta}_b, \boldsymbol{\gamma}_b)$ for $j \in \mathcal{J}_b$. Define the block-removed residual that adds the block's current contribution back,

$$\boldsymbol{r}^{(-b)} := \tilde{\boldsymbol{r}} + \widetilde{\boldsymbol{X}}_b \boldsymbol{\beta}_b. \qquad (45)$$

When $z_b = 0$, the entire block is silenced ($\boldsymbol{\beta}_b = \boldsymbol{0}$ deterministically) and the marginal log-likelihood is simply

$$\log \mathrm{ML}_0 = -1/2\sigma_\varepsilon^2 \, \| \boldsymbol{r}^{(-b)} \|_2^2 + \text{const.} \qquad (46)$$

When $z_b = 1$, the per-SNP cascade in §S1.1.2 is in force, and the marginal likelihood involves a sum over all $2^{k_b}$ possible active-set configurations,

$$p(\boldsymbol{r}^{(-b)} \mid z_b = 1, \cdot) = \sum_{\boldsymbol{\gamma}_b \in \{0,1\}^{k_b}} \Pr(\boldsymbol{\gamma}_b \mid \boldsymbol{\pi}_{\mathcal{J}_b}) \int p(\boldsymbol{r}^{(-b)} \mid \boldsymbol{\beta}_b) \, p(\boldsymbol{\beta}_b \mid \boldsymbol{\gamma}_b) \, \mathrm{d}\boldsymbol{\beta}_b. \qquad (47)$$

For an active set $\mathcal{S} = \{j \in \mathcal{J}_b : \gamma_j = 1\}$, the inner Gaussian integral is tractable by the standard marginalization identity

$$\int \mathcal{N}(\boldsymbol{r}; \widetilde{\boldsymbol{X}}_\mathcal{S} \boldsymbol{\beta}_\mathcal{S}, \sigma_\varepsilon^2 \boldsymbol{I}_n) \, \mathcal{N}(\boldsymbol{\beta}_\mathcal{S}; \boldsymbol{0}, \sigma_\varepsilon^2 \eta_\beta \boldsymbol{I}_{|\mathcal{S}|}) \, \mathrm{d}\boldsymbol{\beta}_\mathcal{S} = \mathcal{N}\big(\boldsymbol{r}; \, \boldsymbol{0}, \, \sigma_\varepsilon^2 \boldsymbol{I}_n + \sigma_\varepsilon^2 \eta_\beta \widetilde{\boldsymbol{X}}_\mathcal{S} \widetilde{\boldsymbol{X}}_\mathcal{S}^\top\big), \qquad (48)$$

but the outer sum over $2^{k_b}$ active-set configurations, each requiring an $|\mathcal{S}| \times |\mathcal{S}|$ Gaussian determinant, is intractable for $k_b \gtrsim 20$.

*S2.4.2 Diagonal approximation within blocks*

We approximate the exact marginal eq. 47 by treating $\widetilde{\boldsymbol{X}}_b^\top \widetilde{\boldsymbol{X}}_b$ as diagonal *for the marginal step only*. Under this approximation the active-set sum factorizes across SNPs, and each SNP contributes its own per-SNP marginal sum $\sum_{\gamma_j \in \{0,1\}} \Pr(\gamma_j)\, p(\boldsymbol{r}_{-j} \mid \gamma_j)$. The $\beta_j$ integration inside each per-SNP contribution is exact under the slab prior — only the cross-SNP independence is approximated. The result is

$$\mathrm{logML}_1 \;=\; -1/2\sigma_\varepsilon^2 \;\|\, \boldsymbol{r}^{(-b)} \,\|_2^2 + \sum_{j \in \mathcal{I}_b} \log\left((1-\pi_j) + \pi_j\, e^{\mathrm{logBF}_j}\right) + \mathrm{const.} \qquad (49)$$

To characterize the approximation error, write $\boldsymbol{Q} = \widetilde{\boldsymbol{X}}_b^\top \widetilde{\boldsymbol{X}}_b \in \mathbb{R}^{k_b \times k_b}$ and decompose $\boldsymbol{Q} = \boldsymbol{D} + \boldsymbol{R}$ where $\boldsymbol{D} = \mathrm{diag}(\boldsymbol{Q})$. The exact marginal log-likelihood involves $\mathrm{logdet}(\eta_\beta \boldsymbol{Q} + \boldsymbol{I})$ over the active SNPs, while the approximation uses $\mathrm{logdet}(\eta_\beta \boldsymbol{D} + \boldsymbol{I}) = \sum_j \log(1 + \eta_\beta Q_{jj})$. Using the second-order Taylor expansion $\mathrm{logdet}(\boldsymbol{I} + \boldsymbol{A}) \approx \mathrm{tr}(\boldsymbol{A}) - 1/2\,\mathrm{tr}(\boldsymbol{A}^2)$ for small $\boldsymbol{A}$ gives the approximation

$$\Delta_{\mathrm{approx}} \;\approx\; 1/2\;\eta_\beta^2 \sum_{i \neq j} \frac{R_{ij}^2}{(1 + \eta_\beta Q_{ii})(1 + \eta_\beta Q_{jj})}. \qquad (50)$$

This bound is smallest when within-block off-diagonal correlations $|R_{ij}|$ are small and increases with stronger residual LD inside a block. In the current sampler, this diagonalization is therefore treated as an approximation for the block-level model-comparison step rather than as an exact marginal likelihood.

The within-block per-SNP scan (§S2.3.1-§S2.3.3) does not use this approximation: LD coupling is retained through the sequential residual update. The diagonal approximation affects only the binary $z_b$ decision, not the per-SNP $\beta_j$ posterior. This two-tier strategy uses the sum of per-SNP marginal evidences in eq. 49 for the coarse block-level signal assessment, then relies on the sequential SNP-level scan to resolve fine-scale posterior effects within active blocks.

*S2.4.3 Sampling $z_b$*

Combining eq. 46 and eq. 49 with the block prior eq. 9,

$$\Pr(z_b = 1 \mid \cdot) \;=\; \frac{\pi_{\mathrm{block}}\, e^{\mathrm{logML}_1}}{\pi_{\mathrm{block}}\, e^{\mathrm{logML}_1} + (1 - \pi_{\mathrm{block}})\, e^{\mathrm{logML}_0}}. \qquad (51)$$

If $z_b$ flips $1 \to 0$, force_block_zero zeros $(\beta_j, \gamma_j)$ for all $j \in \mathcal{I}_b$ and adds the cleared block contribution back into $\tilde{\boldsymbol{r}}$. If $z_b = 1$, gibbs_scan_block_inplace_e13 runs the per-SNP scan of §S2.3.

### S2.5 Variance-scale, block-prior, and annotation-prior updates

After updating the dense component, sparse SNP effects, and block indicators, the sampler refreshes the parameters controlling scale and sparsity. The variance scales determine the allocation of variation among residual noise, sparse effects, and the dense background; the block-prior probability controls the expected number of active LD regions; and the annotation-prior coefficients learn the TSS-distance trend in SNP inclusion.

The remaining unknowns are the variance-scale parameters $(\sigma_\varepsilon^2, \eta_\beta, \eta_g)$, the block-level activation rate $\pi_{\text{block}}$, and the annotation-prior coefficients $(\alpha, \kappa)$. They are updated at the end of each Gibbs sweep conditional on the current latent quantities $(\boldsymbol{\beta}, \boldsymbol{\gamma}, \boldsymbol{z}, \boldsymbol{g})$.

The conjugate hyperpriors in §S1.1.3 give closed-form updates for the variance-scale parameters and the block-prior probability. The annotation-prior coefficients require Metropolis-Hastings because the per-SNP inclusion probability $\pi_j = \sigma(\alpha + \kappa\, a_j)$ enters through a nonconjugate logistic link.

Each variance-scale parameter is updated by retaining its terms in the joint log-density eq. 24 and matching the inverse-gamma kernel. For a positive variable $u$, the $\text{InvGamma}(a, b)$ density has log-kernel $-(a+1)\log u - b/u$; therefore, a conditional log-density of the form "$-A\log u - B/u + \text{const}$" corresponds to $\text{InvGamma}(A-1, B)$. The block-prior probability is handled analogously with the Beta-Bernoulli kernel.

**Residual variance.** Collecting all $\sigma_\varepsilon^2$-dependent terms in eq. 24 — the likelihood normalizer and exponent, the slab normalizer and exponent (which carry the same $\sigma_\varepsilon^2$ that scales the slab variance $\sigma_\varepsilon^2 \eta_\beta$), the polygenic prior normalizer and exponent (likewise), and the $\sigma_\varepsilon^2$ hyperprior — gives

$$\log p(\sigma_\varepsilon^2 \mid \cdot) \stackrel{c}{=} -\left(a_\varepsilon + n + n_\gamma + n_{\text{eff}}/2 + 1\right)\log \sigma_\varepsilon^2 - \frac{1}{\sigma_\varepsilon^2}\left(b_\varepsilon + 1/2\,(\text{SSR} + Q_{\text{slab}} + Q_{\text{poly}})\right), \qquad (52)$$

where $\text{SSR} := \| \tilde{\boldsymbol{y}} - \tilde{\boldsymbol{X}}\boldsymbol{\beta} - \tilde{\boldsymbol{g}} \|_2^2$ is the residual sum of squares in the rotated basis, $Q_{\text{slab}} := \sum_{j:\gamma_j=1} \beta_j^2 / \eta_\beta$ is the slab quadratic form, $Q_{\text{poly}} := \sum_{i:\lambda_i>0} \tilde{g}_i^2 / (\eta_g \lambda_i)$ is the polygenic quadratic form in the rotated basis (using $\boldsymbol{g}^\top \boldsymbol{K}^+ \boldsymbol{g} = \sum_{i:\lambda_i>0} \tilde{g}_i^2 / \lambda_i$), and $n_{\text{eff}} := |\{i : \lambda_i > 10^{-10}\}|$ is the rank-effective dimension. eq. 52 is the kernel of an inverse-gamma distribution. Reading off the parameters,

$$\sigma_\varepsilon^2 \mid \cdot \sim \text{InvGamma}\left(a_\varepsilon + n + n_\gamma + n_{\text{eff}}/2,\; b_\varepsilon + 1/2\,(\text{SSR} + Q_{\text{slab}} + Q_{\text{poly}})\right). \qquad (53)$$

**Slab scale.** Only the slab prior eq. 12 and its hyperprior depend on $\eta_\beta$:

$$\log p(\eta_\beta \mid \cdot) \overset{c}{=} -\left(a_\beta + n_\gamma/2 + 1\right)\log\eta_\beta - \frac{1}{\eta_\beta}\left(b_\beta + 1/2\sigma_\varepsilon^2 \sum_{j:\gamma_j=1} \beta_j^2\right), \qquad (54)$$

which gives

$$\eta_\beta \mid \cdot \sim \text{InvGamma}\left(a_\beta + n_\gamma/2,\ b_\beta + 1/2\sigma_\varepsilon^2 \sum_{j:\gamma_j=1} \beta_j^2\right). \qquad (55)$$

**Polygenic scale.** The rotated polygenic prior contributes $-n_{\text{eff}}/2\log\eta_g$ from the determinant of $\sigma_\varepsilon^2\eta_g\mathbf{\Lambda}$ restricted to the positive eigenvalues, and $\sum_i \tilde{g}_i^2/(\sigma_\varepsilon^2\eta_g\lambda_i)$ from the quadratic. Combined with the hyperprior,

$$\log p(\eta_g \mid \cdot) \overset{c}{=} -\left(a_g + n_{\text{eff}}/2 + 1\right)\log\eta_g - \frac{1}{\eta_g}\left(b_g + 1/2\sigma_\varepsilon^2 \sum_{i:\lambda_i>0} \tilde{g}_i^2/\lambda_i\right), \qquad (56)$$

hence

$$\eta_g \mid \cdot \sim \text{InvGamma}\left(a_g + n_{\text{eff}}/2,\ b_g + 1/2\sigma_\varepsilon^2 \sum_{i:\lambda_i>0} \tilde{g}_i^2/\lambda_i\right). \qquad (57)$$

**Block-prior probability.** Extracting the $\pi_{\text{block}}$-dependent terms from eq. 24 — the block factor eq. 9 times the Beta(1,1) hyperprior, which contributes nothing — leaves the kernel

$$\log p(\pi_{\text{block}} \mid \boldsymbol{z}) \overset{c}{=} \left(\sum_b z_b\right)\log\pi_{\text{block}} + \left(B - \sum_b z_b\right)\log(1-\pi_{\text{block}}), \qquad (58)$$

which is the kernel of $\text{Beta}(1 + \sum_b z_b,\ 1 + B - \sum_b z_b)$. Hence

$$\pi_{\text{block}} \mid \boldsymbol{z} \sim \text{Beta}\left(1 + \sum_b z_b,\ 1 + B - \sum_b z_b\right), \qquad (59)$$

clipped numerically to $[1/\max(B,10), 1 - 1/\max(B,10)]$ to prevent zero-probability draws when $\sum_b z_b \in \{0, B\}$. Implementation: variance and $\pi_{\text{block}}$ updates are inlined in bsbslmm/bsbslmm_mcmc.py::bsbslmm_gibbs (eq. 53 at lines 233–245, eq. 55 at 247–256, eq. 57 at 258–263, eq. 59 at 273–279).

*S2.5.1 Random-walk Metropolis-Hastings on* $(\alpha, \kappa)$

The annotation-prior coefficients are the only conditional in the model without a closed-form Gibbs update. The per-SNP inclusion probability $\pi_j = \sigma(\alpha + \kappa\, a_j)$ enters the prior on $\gamma_j$ through a nonconjugate logistic link. Because the parameter space is two-dimensional, we use random-walk Metropolis-Hastings. Conditioning on the active-block subset $\mathcal{A} = \{j: z_{b(j)} = 1\}$, which contains the only SNPs whose $\gamma_j$ depends on the annotation-prior coefficients, the conditional log-posterior is, from eq. 24,

$$\log p(\alpha, \kappa \mid \boldsymbol{\gamma}, \boldsymbol{z}) \stackrel{c}{=} \sum_{j \in \mathcal{A}} [\gamma_j \log \pi_j + (1 - \gamma_j) \log(1 - \pi_j)] - \frac{(\alpha - \alpha_0)^2}{2\tau_\alpha^2} - \frac{\kappa^2}{2\tau_\kappa^2}. \tag{60}$$

Using the logistic identities $\log\sigma(x) = x - \log(1 + e^x)$ and $\log(1 - \sigma(x)) = -\log(1 + e^x)$ with $\ell_j = \alpha + \kappa\, a_j$, the Bernoulli sum collapses to

$$\sum_{j \in \mathcal{A}} [\gamma_j \log \pi_j + (1 - \gamma_j) \log(1 - \pi_j)] = \sum_{j \in \mathcal{A}} [\gamma_j \ell_j - \log(1 + e^{\ell_j})], \tag{61}$$

so that

$$\log p(\alpha, \kappa \mid \boldsymbol{\gamma}, \boldsymbol{z}) \stackrel{c}{=} \sum_{j \in \mathcal{A}} [\gamma_j \ell_j - \log(1 + e^{\ell_j})] - \frac{(\alpha - \alpha_0)^2}{2\tau_\alpha^2} - \frac{\kappa^2}{2\tau_\kappa^2}. \tag{62}$$

This is a logistic-regression posterior with normal priors. We update it via a 2-D random-walk Metropolis-Hastings step. At each MH step we propose

$$(\alpha', \kappa') = (\alpha, \kappa) + (\delta_\alpha\, \xi_1,\, \delta_\kappa\, \xi_2), \qquad \xi_1, \xi_2 \sim \mathcal{N}(0,1), \tag{63}$$

with step sizes $(\delta_\alpha, \delta_\kappa) = (0.3, 0.5)$, compute the log-posterior difference $\Delta = \log p(\alpha', \kappa' \mid \cdot) - \log p(\alpha, \kappa \mid \cdot)$ via eq. 62, and accept with probability $\min(1, e^\Delta)$. We perform four such MH steps per Gibbs sweep to reduce within-sweep autocorrelation. Implementation: bsbslmm/bsbslmm_mcmc.py::_update_alpha_beta_mh.

### S2.6 Pseudocode

The following algorithm summarizes one Gibbs sweep of bsBSLMM. The eigendecomposition of the GRM and the rotation of the design matrix are computed once at startup; the loop body executes the six conditional updates derived in §S2 and stores thinned samples. Equations cited in the comments refer to this document.

```
1  input:  X (n×p), y (n), block_indices, a_per_snp (p), n_iter, n_burn, n_thin, π_init=0.05
2  output: β̄, σ(β), q̂_{0.025}(β), q̂_{0.975}(β), γ̄, z̄, hyperparameter posteriors
3
4  # precomputation (once per gene)
5  Λ, U ← eigendecompose K = (1/p) X X^T          # O(n³); Eq. (S25)
6  X̃  ← U^T X                                    # O(n²p)
7  ỹ  ← U^T y                                    # O(n²)
8
9  # initialization
10 β ← 0, γ ← 0, z ← 0, g̃ ← 0
11 σ_ε², η_β, η_g, π_block ← empirical defaults
12 α ← logit(π_init);  κ ← 0
```

```
13
14 # MCMC
15 for it = 1..n_iter:
16    ℓ_j ← α + κ · a_j (∀j);  π_j ← σ(ℓ_j)          # Eq. (S13)
17
18    # block + per-SNP cis-effect updates (§S2.3–§S2.4)
19    r̃ ← ỹ − X̃ β − g̃
20    for b = 1..B:
21       r⁽⁻ᵇ⁾ ← r̃ + X̃_b β_b                       # Eq. (S45)
22       log_ML0 ← −‖r⁽⁻ᵇ⁾‖² / (2σ_ε²)               # Eq. (S46)
23       log_ML1 ← log_ML0
24              + Σ_{j∈I_b} log[(1−π_j) + π_j · BF_j]   # Eq. (S49)
25       sample z_b from Eq. (S51)
26       if z_b = 1:
27          for j ∈ I_b (sequential):
28             r_{−j}    ← r̃ + X̃_{:,j} β_j       # Eq. (S32)
29             v_j, μ_j  ← Eq. (S36);  BF_j ← Eq. (S42)
30             sample γ_j ~ Bernoulli(Eq. (S43))
31             if γ_j = 1:
32                sample β_j ~ N(μ_j, v_j)
33             else:
34                β_j ← 0
35             r̃ ← r_{−j} − X̃_{:,j} β_j
36       else:
37          r̃ ← r⁽⁻ᵇ⁾;  β_b ← 0;  γ_b ← 0
38
39    # polygenic update (§S2.2.2)
40    for i = 1..n:
41       sample g̃_i from Eq. (S31)
42
43    # variance-scale parameters (§S2.5)
44    σ_ε² ← Eq. (S53)
45    η_β  ← Eq. (S55)
46    η_g  ← Eq. (S57)
47
48    # annotation prior (§S2.5) — implemented before π_block in
49    # bsbslmm/bsbslmm_mcmc.py (the within-sweep order is MH-then-Beta;
50    # this is a Gibbs-scan order choice with no effect on the stationary
51    # distribution since (α, κ) and π_block are conditionally independent
52    # given (β, γ, z) within a sweep)
53    repeat 4 times:
54       propose (α', κ') ~ Eq. (S73)
55       accept with min(1, exp(Δ)) using Eq. (S72)
56
57    # block-inclusion prior (§S2.5)
58    π_block ← Eq. (S59)
59
60    # store sample
61    if it ≥ n_burn and (it − n_burn) mod n_thin = 0:
62       record(β, γ, z, α, κ, π_block, σ_ε², η_β, η_g)
63
64 return posterior moments and quantiles              # Eq. (S74)
```

### S2.7 Runtime equation manifest

The 18 equations listed below are the closed-form update rules actually evaluated at runtime by bsbslmm/bsbslmm_mcmc.py::bsbslmm_gibbs and the inner Numba kernel bsbslmm/_mcmc_kernel.py. Every other display equation in §S1-S2 is a model-specification statement (likelihood, prior, joint factorization) or a derivation step (rotation lemma, completing the square, Bayes-factor reduction, conditional log-density assembly) whose role is to justify one of the terminal updates below. They are retained for transparency but are never invoked during sampling.

| Stage | Eq. | Quantity computed | Frequency |
|---|---|---|---|
| One-time setup | eq. 25 | eigendecomposition $\boldsymbol{K} = \boldsymbol{U\Lambda U}^\top$ | once per gene |
| Annotation prior | eq. 10 | $\pi_j = \sigma(\alpha + \kappa\, a_j)$ | once per sweep |
| Block sweep | eq. 45 | residual restore $\boldsymbol{r}^{(-b)} = \boldsymbol{r} + \widetilde{\boldsymbol{X}}_b \boldsymbol{\beta}_b$ | per block |
| | eq. 46 | $\mathrm{logML}_0 = -\lVert \boldsymbol{r}^{(-b)} \rVert_2^2 / (2\sigma_\varepsilon^2)$ | per block |
| | eq. 49 | $\sum_{j \in \mathcal{J}_b} \log\left((1-\pi_j) + \pi_j \cdot \mathrm{BF}_j\right)$ | per block |
| | eq. 51 | sample $z_b \sim$ Bernoulli | per block |
| Per-SNP sweep (active blocks) | eq. 32 | residual restore $\boldsymbol{r}_{-j} = \boldsymbol{r} + \widetilde{\boldsymbol{X}}_{:,j} \beta_j$ | per active SNP |
| | eq. 36 | slab posterior $(v_j, \mu_j)$ and $\beta_j$ draw | per active SNP |
| | eq. 42 | Bayes factor $\mathrm{BF}_j$ | per active SNP |
| | eq. 43 | sample $\gamma_j \sim$ Bernoulli | per active SNP |
| Polygenic background | eq. 31 | sample $\tilde{g}_i$ in rotated basis | $n_{\mathrm{eff}}$ scalars/sweep |
| Variance / mixing | eq. 53 | $\sigma_\varepsilon^2$ posterior InvGamma draw | once per sweep |
| | eq. 55 | $\eta_\beta$ posterior InvGamma draw | once per sweep |
| | eq. 57 | $\eta_g$ posterior InvGamma draw | once per sweep |
| | eq. 59 | $\pi_{\mathrm{block}}$ posterior Beta draw | once per sweep |
| Annotation MH | eq. 63 | propose $(\alpha', \kappa')$ | 4× per sweep |
| | eq. 62 | accept via log-posterior difference | 4× per sweep |
| Postprocess | eq. 64 | posterior means, SDs, and 2.5 %/97.5 % quantiles | after warmup |

In total, 18 closed-form rules drive the chain. The remaining display equations in §S1-S2 partition into:

- **Model specification** (likelihood, priors, joint density): eq. 1-eq. 24 excluding the runtime rules above.
- **Derivations** that produce the runtime rules: the rotation lemma and its consequences (§S2.2) leading to eq. 31; completing-the-square steps (§S2.3.1) leading to eq. 36; Bayes-factor reduction (§S2.3.2) leading to eq. 42; active-set marginalization (§S2.4) leading to eq. 51; inverse-gamma kernel matching (§S2.5) leading to eq. 53/eq. 55/eq. 57; posterior moment definitions (§S3.1) leading to eq. 64; and the predictor decomposition (§S3.2) leading to eq. 72-eq. 73.

Readers reproducing the sampler from scratch need only implement the 18 update rules above in the order shown by the pseudocode in §S2.6.

### S2.8 Computational complexity

Per gene, the eigendecomposition of $\boldsymbol{K}$ costs $O(n^3)$, the rotation $\boldsymbol{U}^\top \boldsymbol{X}$ costs $O(n^2 p)$, and rotating $\boldsymbol{y}$ costs $O(n^2)$. For chr20 cis windows ($n \approx 280$, $p \approx 3000$), $n^3 \approx 2.2 \times 10^7$ and $n^2 p \approx$ $2.4 \times 10^8$, so the rotation dominates the one-time precomputation and is amortized over the Gibbs sweeps.

Per Gibbs sweep, the block on/off marginal step costs $O(np)$ aggregate (each $\log\mathrm{BF}_j$ requires the inner product $\tilde{\boldsymbol{X}}_{:,j}^{\top}\boldsymbol{r}^{(-b)}$ at $O(n)$), the per-SNP scan over active blocks costs $O(n|\mathcal{A}|)$ (each SNP performs an $O(n)$ residual update), the polygenic update costs $O(n)$ (one scalar Gaussian per eigencomponent), the variance updates cost $O(n + p)$ aggregate, the four MH steps cost $O(|\mathcal{A}|)$ each, and the Beta-Bernoulli update costs $O(B)$. The per-sweep cost is dominated by the $O(np)$ block-scan and per-SNP-scan terms; aggregated over $T = n_{\mathrm{iter}}$ iterations, the total per-gene cost is $O(n^3 + n^2 p + Tnp)$.

For chr20 ($n \approx 280, p \approx 3000, T = 5{,}000$), the dominant term is $Tnp \approx 4 \times 10^9$, about $30 \times$ the eigendecomposition cost. In the dedicated single-thread speed benchmark (results/benchmark_G/analysis_speed/speed_chr20.csv), the Numba JIT kernel (bsbslmm/_mcmc_kernel.py) gives a median bsBSLMM full-fit time of 55.6 seconds per gene, corresponding to roughly 9.4 minutes for the 606-gene chromosome at $J = 60$ slots under ideal dispatch.

The production schedule is $T = 5{,}000$, $n_{\mathrm{burn}} = 2{,}500$, $n_{\mathrm{thin}} = 25$, giving $M = 100$ thinned post-burn samples. The simulation experiments (§3.1 of the main text) use a smaller schedule ($T = 1{,}500$, $n_{\mathrm{burn}} = 300$, $n_{\mathrm{thin}} = 12$) for tractability across the full method × scenario × gene grid; all real-data fits including the genome-wide deployment use the full $5{,}000/2{,}500/25$ schedule.

The thinning factor of 25 was selected to reduce post-burn autocorrelation while keeping the per-gene output compact. The burn-in of 2,500 sweeps discards 50% of the production schedule before posterior summaries are computed.

## S3. Posterior summaries, prediction, and implementation

The previous section derives the conditional updates used by the sampler. This section specifies the retained MCMC summaries, the prediction weights used for held-out evaluation, and the implementation footprint of the released codebase.

The MCMC chain returns latent configurations $\{(\boldsymbol{\beta}^{(m)}, \boldsymbol{\gamma}^{(m)}, \boldsymbol{z}^{(m)}, \ldots)\}_{m=1}^{M}$. Four per-SNP summaries are retained: the posterior mean $\hat{\beta}_j$, the posterior standard deviation $\widehat{\mathrm{SD}}_j$, and the empirical 2.5 % and 97.5 % quantiles $[\hat{q}_{j,0.025}, \hat{q}_{j,0.975}]$. The posterior mean $\hat{\beta}_j$ is used for held-out prediction (§S3.2) and exported as a PredictDB-format weight compatible with PrediXcan [1] and SPrediXcan [2]. Downstream TWAS analyses use posterior mean weights $\widehat{\boldsymbol{\beta}}$ together with the in-sample LD covariance computed from training-cohort genotypes. Empirical posterior intervals define active SNPs and support case-study uncertainty visualization. A SNP is called credibly active if its 95% empirical posterior interval excludes zero. Empirical quantiles are used because the marginal posterior of $\beta_j$ is a spike-and-slab mixture.

### S3.1 Empirical posterior intervals for sparse effects

Let $\{\boldsymbol{\beta}^{(m)}\}_{m=1}^{M}$ denote the post-burn, post-thin samples, where $M = (n_{\text{iter}} - n_{\text{burn}})/n_{\text{thin}} = 100$ under the production schedule of $5{,}000/2{,}500/25$. For each SNP $j$ we compute the posterior mean, posterior standard deviation, and the empirical $0.025$ and $0.975$ quantiles,

$$\hat{\beta}_j = \frac{1}{M}\sum_{m=1}^{M} \beta_j^{(m)}, \qquad \widehat{\mathrm{SD}}_j = \sqrt{\frac{1}{M}\sum_{m=1}^{M} (\beta_j^{(m)} - \hat{\beta}_j)^2}, \qquad \hat{q}_{j,p} = \mathrm{Quantile}_p\left(\{\beta_j^{(m)}\}_{m=1}^{M}\right), \qquad (64)$$

with $p \in \{0.025, 0.975\}$. These are saved as beta, beta_std, beta_q025, beta_q975 in beta.npz (scripts/run_bsbslmm_chunk.py).

The marginal posterior of $\beta_j$ is a mixture: with probability $\Pr(\gamma_j = 0 \mid \boldsymbol{y})$ a point mass at zero, and with probability $\tilde{\pi}_j := \Pr(\gamma_j = 1 \mid \boldsymbol{y})$ a Gaussian centred at the conditional posterior mean $\mu_j^*$ with conditional posterior variance $v_j^*$,

$$p(\beta_j \mid \boldsymbol{y}) = (1 - \tilde{\pi}_j)\,\delta_0(\beta_j) + \tilde{\pi}_j\,\mathcal{N}(\beta_j; \mu_j^*, v_j^*). \qquad (65)$$

Under eq. 65, the unconditional moments are

$$\hat{\beta}_j = \tilde{\pi}_j\,\mu_j^*, \qquad \widehat{\mathrm{SD}}_j^2 = \tilde{\pi}_j(1 - \tilde{\pi}_j)\,(\mu_j^*)^2 + \tilde{\pi}_j\,v_j^*. \qquad (66)$$

A symmetric Gaussian interval $[\hat{\beta}_j - 1.96\,\widehat{\mathrm{SD}}_j, \hat{\beta}_j + 1.96\,\widehat{\mathrm{SD}}_j]$ is inappropriate whenever $\tilde{\pi}_j \notin \{0,1\}$ because the marginal posterior contains both a point mass and a slab component. Under MCMC ergodicity, the empirical quantile $\hat{q}_{j,p}$ consistently estimates the corresponding posterior quantile. Thus $[\hat{q}_{j,0.025}, \hat{q}_{j,0.975}]$ directly represents the central 95 % posterior interval for the mixture posterior in eq. 65.

### S3.2 Held-out prediction

The posterior-predictive distribution of test-cohort expression is Gaussian, with mean obtained by conditioning the test-cohort polygenic effect on the training-cohort residual. This is the standard linear-mixed-model BLUP construction: train and test polygenic effects are jointly Gaussian under kernel $\boldsymbol{K}$, and conditioning yields a kernel-ridge adjustment to the sparse cis-effect predictor.

#### *S3.2.1 Mixed-model BLUP for held-out cohorts*

For a held-out test cohort $\boldsymbol{X}^{\text{test}} \in \mathbb{R}^{n_{\text{test}} \times p}$ and the training data $(\boldsymbol{X}^{\text{train}}, \boldsymbol{y}^{\text{train}})$ with posterior-mean cis-effect $\widehat{\boldsymbol{\beta}}$, the posterior-predictive mean of test expression is a Gaussian-conditioning mixed-model BLUP. We derive it directly from the joint Gaussian distribution.

The training-cohort residual is $\boldsymbol{e} := \boldsymbol{y}^{\text{train}} - \boldsymbol{X}^{\text{train}}\widehat{\boldsymbol{\beta}} = \boldsymbol{g}^{\text{train}} + \boldsymbol{\varepsilon}^{\text{train}}$, where

$$\boldsymbol{g}^{\text{train}} \sim \mathcal{N}(\boldsymbol{0}, \sigma_\varepsilon^2 \eta_g \boldsymbol{K}^{\text{train}}), \qquad \boldsymbol{\varepsilon}^{\text{train}} \sim \mathcal{N}(\boldsymbol{0}, \sigma_\varepsilon^2 \boldsymbol{I}_n). \qquad (67)$$

The joint distribution of $(\boldsymbol{g}^{\text{train}}, \boldsymbol{g}^{\text{test}})$ is multivariate Gaussian with cross-covariance $\sigma_\varepsilon^2 \eta_g \boldsymbol{K}^{\text{test,train}}$, where $\boldsymbol{K}^{\text{test,train}} = 1/p\, \boldsymbol{X}^{\text{test}}(\boldsymbol{X}^{\text{train}})^\top$. By the standard Gaussian conditioning formula, the conditional distribution of $\boldsymbol{g}^{\text{test}}$ given the training residual $\boldsymbol{e}$ has mean

$$\mathbb{E}[\boldsymbol{g}^{\text{test}} \mid \boldsymbol{e}] = \text{Cov}(\boldsymbol{g}^{\text{test}}, \boldsymbol{e})\, \text{Cov}(\boldsymbol{e})^{-1}\, \boldsymbol{e}. \qquad (68)$$

Computing each piece separately, the cross-covariance is

$$\text{Cov}(\boldsymbol{g}^{\text{test}}, \boldsymbol{e}) = \text{Cov}(\boldsymbol{g}^{\text{test}}, \boldsymbol{g}^{\text{train}}) = \sigma_\varepsilon^2 \eta_g \boldsymbol{K}^{\text{test,train}}, \qquad (69)$$

and the marginal covariance of the training residual is

$$\text{Cov}(\boldsymbol{e}) = \text{Cov}(\boldsymbol{g}^{\text{train}}) + \text{Cov}(\boldsymbol{\varepsilon}^{\text{train}}) = \sigma_\varepsilon^2 \eta_g \boldsymbol{K}^{\text{train}} + \sigma_\varepsilon^2 \boldsymbol{I}_n$$
$$= \sigma_\varepsilon^2\left(\eta_g \boldsymbol{K}^{\text{train}} + \boldsymbol{I}_n\right). \qquad (70)$$

Substituting eq. 69 and eq. 70 into eq. 68 cancels the $\sigma_\varepsilon^2$ factor:

$$\mathbb{E}[\boldsymbol{g}^{\text{test}} \mid \boldsymbol{e}] = \eta_g \boldsymbol{K}^{\text{test,train}}(\eta_g \boldsymbol{K}^{\text{train}} + \boldsymbol{I}_n)^{-1}\boldsymbol{e}. \qquad (71)$$

Adding the cis-effect contribution $\boldsymbol{X}^{\text{test}}\widehat{\boldsymbol{\beta}}$ and substituting the posterior mean $\hat{\eta}_g$ for $\eta_g$ gives the predictor

$$\widehat{\boldsymbol{y}}^{\text{test}} = \boldsymbol{X}^{\text{test}}\widehat{\boldsymbol{\beta}} + \hat{\eta}_g \boldsymbol{K}^{\text{test,train}}(\hat{\eta}_g \boldsymbol{K}^{\text{train}} + \boldsymbol{I}_n)^{-1}\left(\boldsymbol{y}^{\text{train}} - \boldsymbol{X}^{\text{train}}\widehat{\boldsymbol{\beta}}\right). \qquad (72)$$

This is the Gaussian-conditioning predictor corresponding to Eq. (5) of the main manuscript. Implementation: bsbslmm/bslmm_mcmc.py::predict_bslmm, called by bsbslmm/bsbslmm_mcmc.py::predict_bsbslmm.

The first term in eq. 72, $\boldsymbol{X}^{\text{test}}\widehat{\boldsymbol{\beta}}$, is the transferable cis-genetic component because it depends only on test genotypes and training-derived per-SNP weights. The second term is an individual-level polygenic adjustment determined by train-test genetic similarity ($\boldsymbol{K}^{\text{test,train}}$) and the estimated polygenic-background variance ratio ($\hat{\eta}_g$).

In the production evaluation pipeline these two terms are applied in different contexts. The within-CV evaluation called from scripts/run_bsbslmm_chunk.py uses the full eq. 72 predictor (implemented as bsbslmm/bsbslmm_mcmc.py::predict_bsbslmm, which composes bslmm_mcmc.py::predict_bslmm); this is what generates the cv_r values reported per fold during training. The individual-level held-out GEUVADIS evaluation in §3.2 of the main text uses the sparse-cis-only specialization

$$\widehat{\boldsymbol{y}}^{\text{test}} = \boldsymbol{X}^{\text{test}}\,\widehat{\boldsymbol{\beta}}, \qquad (73)$$

implemented at scripts/inference_bsbslmm.py. The same X_test_std @ beta form is used by every baseline: BSLMM, LASSO, BLUP, TIGAR_EN in the TIGAR framework with elastic-net regularization, and TIGAR_DPR using latent Dirichlet-process regression in the TIGAR framework. Thus, the held-out paired head-to-head comparisons use a common individual-level predictor form.

In SPrediXcan-style summary-statistics TWAS, only the transferable weight vector $\widehat{\boldsymbol{\beta}}$ and the in-sample LD covariance from the training cohort are used, because individual-level GWAS genotypes and the train–test cross-kernel are not available. ### S3.3 Implementation footprint

The sampler orchestrator is bsbslmm/bsbslmm_mcmc.py, which implements the Gibbs schedule of §S2.6. The Numba-JIT kernel bsbslmm/_mcmc_kernel.py provides the per-SNP Gibbs scan (§S2.3.1-S2.3.3) and the block-marginal log-evidence (§S2.4). The eigendecomposition precomputation, the rotated polygenic sampler (§S2.2.2), and the mixed-model predictor (§S3.2) reside in bsbslmm/bslmm_mcmc.py. The per-chunk runtime entry point is scripts/run_bsbslmm_chunk.py, which is dispatched as a Slurm array job by job/bsbslmm_array.batch and finalized by job/bsbslmm_finalize_chr.batch for test-cohort inference and per-chromosome summaries.

# Part II. Supplementary Analyses

This section reports extended benchmark tables, secondary analyses, runtime measurements, and source-data mappings supporting the main manuscript.

## S1. Data, preprocessing, and benchmark design

The primary benchmark used GEUVADIS European-ancestry lymphoblastoid cell line (LCL) expression from Battle et al. [3] and Lappalainen et al. [4], together with 1000 Genomes Project Phase 3 genotypes [5] on Genome Reference Consortium Human Build 38 (GRCh38) [6]. The European subset included four 1000 Genomes population codes: Utah Residents with Northern and Western European Ancestry (CEU), Finnish in Finland (FIN), British in England and Scotland (GBR), and Toscani in Italia (TSI). Gene expression was normalized and adjusted for hidden confounders following the Genotype-Tissue Expression (GTEx)-style pipeline from the GTEx pilot analysis [7] and GTEx v8 atlas [8]. For each gene, cis single-nucleotide polymorphisms (SNPs) were selected from a 1 Mb window around the transcription start site (TSS).

The primary evaluation set contained 23,098 autosomal genes. Models were evaluated by 5-fold cross-validation and by held-out prediction. The main manuscript applies both prediction filters:

$$\text{cv-}r^2 \geq 0.01 \quad \text{and} \quad \text{test_}r > 0.4. \qquad (74)$$

At the GEUVADIS held-out test size of approximately 40 individuals, $\text{test_}r > 0.4$ corresponds to an approximate two-sided correlation-test threshold of $p \approx 0.011$. The same nominal threshold was used to define matched-power held-out thresholds for the Louisiana Osteoporosis Study (LOS): $r > 0.35$ in the Caucasian American cohort and $r > 0.43$ in the African American cohort.

## S2. Comparative methods and GEUVADIS benchmark details

The benchmark contrasts six prediction methods that span sparse penalised regression, dense polygenic shrinkage, and flexible Bayesian transcriptome imputation. Supplementary Table S1 lists each method together with its model class and the role it plays in the comparison; this is the comparator panel referenced throughout the main-text Results and the remainder of this section.

**Supplementary Table S1.** Genome-wide prediction methods used in the GEUVADIS benchmark.

| Method | Model class | Role in comparison |
|---|---|---|
| LASSO | Least absolute shrinkage and selection operator regression | Standard sparse baseline |
| BLUP | Best linear unbiased prediction / dense linear mixed model | Polygenic baseline |
| BSLMM | Bayesian sparse linear mixed model | Primary Bayesian baseline |
| TIGAR EN | Transcriptome-Integrated Genetic Association Resource elastic net | External imputation baseline |
| TIGAR DPR | Transcriptome-Integrated Genetic Association Resource Dirichlet-process regression | Flexible Bayesian transcriptome-imputation baseline |
| bsBSLMM | Block-sparse BSLMM with TSS prior | Proposed method |

MCMC schedules were matched for BSLMM and bsBSLMM in the genome-wide benchmark: primary runs used $n_{\text{iter}} = 5{,}000$, $n_{\text{burn}} = 2{,}500$, and thinning interval 25 for both methods. The main manuscript reports the number of genes passing both prediction filters and paired held-out comparisons. To complement those counts with a screening-stage view, Supplementary Table S2 lists, for each method on the 23,098-gene autosomal panel, the CV-only pass count, the count passing both filters, and the retention ratio of the latter among the former. The pass-both-filters column reproduces the Table 1 totals from the main text; the CV-only column isolates the screening filter before the held-out-$r$ requirement is applied, exposing how aggressively each method is downweighted by held-out testing.

**Supplementary Table S2.** GEUVADIS CV-only and held-out-filtered yield across all autosomes.

| Method | CV-only pass | Pass both filters | Retention among CV-pass genes |
|---|---|---|---|
| LASSO | 8,387 | 3,164 | 37.7% |
| BLUP | 9,364 | 2,194 | 23.4% |
| BSLMM | 11,750 | 3,130 | 26.6% |

| Method | CV-only pass | Pass both filters | Retention among CV-pass genes |
|---|---|---|---|
| TIGAR EN | 11,121 | 3,129 | 28.1% |
| TIGAR DPR | 19,142 | 3,239 | 16.9% |
| bsBSLMM | 12,236 | 3,518 | 28.8% |

Under the CV-only screen, bsBSLMM had 6,657 wins among 10,890 paired CV-passing genes (61.1%), mean $\Delta r = +0.018$, and one-sided Wilcoxon $p = 8.7 \times 10^{-132}$. This result is consistent with the main prediction-filter analysis, which additionally requires held-out $r > 0.4$. Supplementary Figure S1 plots the same per-method counts broken down by autosome, so the pass-both-filters totals in Supplementary Table S2's middle column can be inspected for chromosomes that disproportionately drive the gap between bsBSLMM and the comparators.

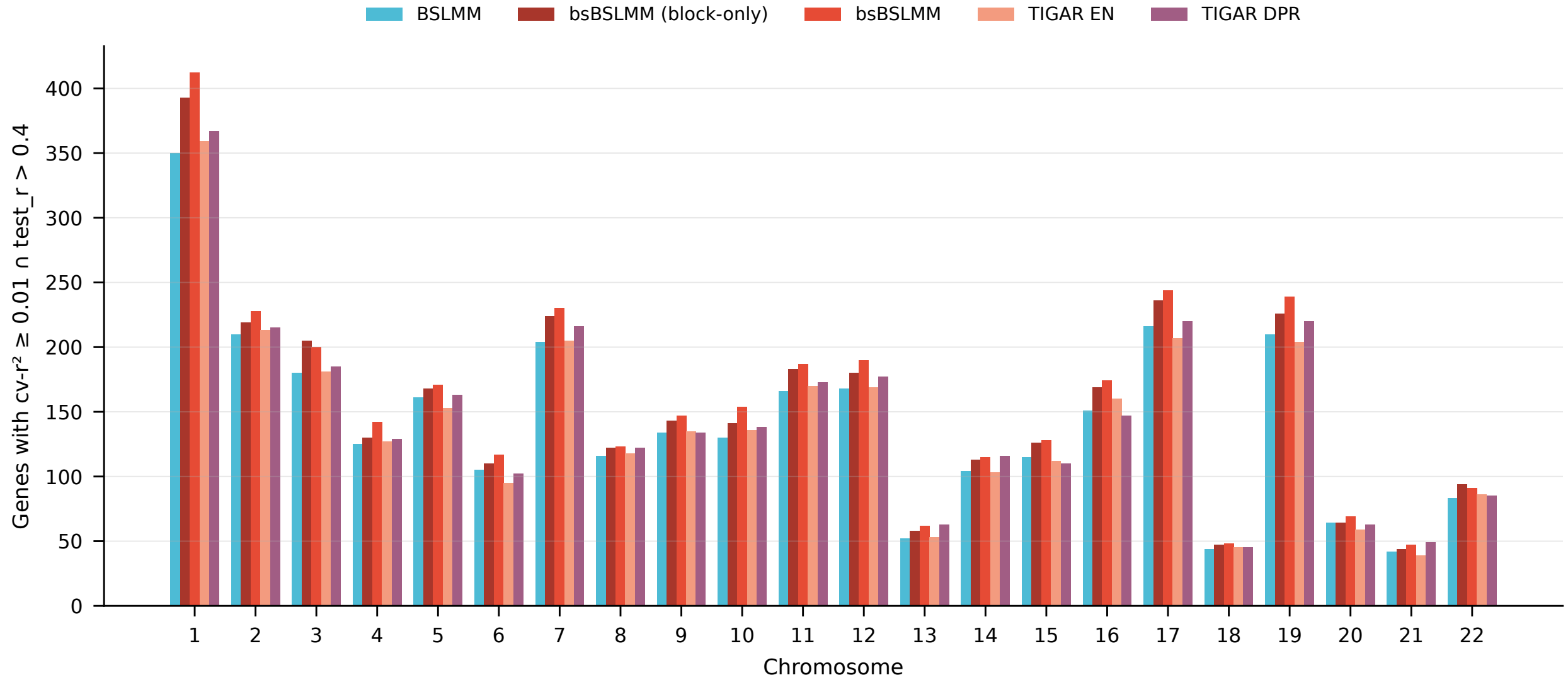

**Supplementary Figure S1.** Per-autosome counts of genes passing both GEUVADIS prediction filters (CV-$r^2 \geq 0.01$ and test-$r > 0.4$). Source figure: manuscript/figures/Fig_S1_genomewide.pdf.

## S2.1 Per-chromosome paired comparison

The genome-wide paired Wilcoxon test reported in Table 1 collapses across 22 autosomes and can be driven by a few large-yield chromosomes. Supplementary Table S3 unwinds that contrast and reports, for each autosome separately, the number of shared genes passing both prediction filters, the bsBSLMM win count and rate, the mean held-out $\Delta r$, and the one-sided Wilcoxon $p$. Every chromosome favours bsBSLMM (win rate 59-80%, mean $\Delta r$ between $+0.005$ and $+0.036$, all $p < 0.05$), so the genome-wide signal is not localised to a single segment of the genome.

**Supplementary Table S3.** Per-chromosome bsBSLMM vs BSLMM head-to-head comparison among genes passing both prediction filters.

| Chromosome | Shared genes | bsBSLMM wins | Win rate | Mean $\Delta r$ | Wilcoxon $p$ |
|---|---|---|---|---|---|
| 1 | 334 | 233 | 69.8% | +0.0222 | $8.1 \times 10^{-17}$ |
| 2 | 191 | 124 | 64.9% | +0.0151 | $4.6 \times 10^{-7}$ |
| 3 | 165 | 117 | 70.9% | +0.0210 | $1.7 \times 10^{-9}$ |
| 4 | 118 | 85 | 72.0% | +0.0196 | $7.3 \times 10^{-6}$ |
| 5 | 145 | 99 | 68.3% | +0.0203 | $1.1 \times 10^{-7}$ |
| 6 | 97 | 68 | 70.1% | +0.0179 | $2.0 \times 10^{-5}$ |
| 7 | 196 | 129 | 65.8% | +0.0166 | $4.2 \times 10^{-8}$ |
| 8 | 107 | 76 | 71.0% | +0.0229 | $4.2 \times 10^{-6}$ |
| 9 | 127 | 85 | 66.9% | +0.0163 | $3.0 \times 10^{-5}$ |
| 10 | 120 | 77 | 64.2% | +0.0141 | $6.1 \times 10^{-4}$ |
| 11 | 156 | 108 | 69.2% | +0.0228 | $1.8 \times 10^{-9}$ |
| 12 | 160 | 112 | 70.0% | +0.0205 | $5.1 \times 10^{-8}$ |
| 13 | 48 | 37 | 77.1% | +0.0313 | $3.4 \times 10^{-5}$ |
| 14 | 96 | 59 | 61.5% | +0.0140 | $3.2 \times 10^{-3}$ |
| 15 | 109 | 80 | 73.4% | +0.0167 | $2.2 \times 10^{-6}$ |
| 16 | 143 | 104 | 72.7% | +0.0287 | $2.0 \times 10^{-10}$ |
| 17 | 203 | 139 | 68.5% | +0.0164 | $1.6 \times 10^{-9}$ |
| 18 | 39 | 24 | 61.5% | +0.0161 | $1.5 \times 10^{-2}$ |
| 19 | 197 | 127 | 64.5% | +0.0160 | $6.8 \times 10^{-6}$ |
| 20 | 60 | 38 | 63.3% | +0.0053 | $1.9 \times 10^{-2}$ |
| 21 | 39 | 31 | 79.5% | +0.0355 | $5.3 \times 10^{-6}$ |
| 22 | 78 | 46 | 59.0% | +0.0081 | $1.3 \times 10^{-2}$ |
| **All** | **2,928** | **1,998** | **68.2%** | **+0.0189** | $6.85 \times 10^{-113}$ |

## S2.2 Stratified paired comparison

A further concern is that bsBSLMM's advantage might be confined to a particular slice of cis architecture — for instance, genes with many cis-SNPs, or genes whose blocks have unusually strong within-block LD. Supplementary Table S4 partitions the 2,928 shared genes passing both prediction filters into tertiles along four such axes (mean within-block $|r|$, number of cis-SNPs, number of LD blocks, and gene-window size) and reports bsBSLMM's win rate, mean $\Delta r$, and Wilcoxon $p$ in each tertile. The advantage is consistent across all four stratifications: every tertile shows a 65-71% win rate and a mean $\Delta r$ between $+0.014$ and $+0.029$, with $p < 10^{-30}$ throughout, indicating that the improvement is not confined to a single signal-strength regime.

**Supplementary Table S4.** Stratified bsBSLMM vs BSLMM paired comparison by cis-window and LD summaries.

| Stratum | T1 (low) | T2 (mid) | T3 (high) |
|---|---|---|---|
| Within-block $\|r\|$ | 67.3%, $\Delta = +0.0181$ , $p = 4.7 \times 10^{-37}$ | 68.8%, $\Delta = +0.0206$ , $p = 2.8 \times 10^{-43}$ | 68.6%, $\Delta = +0.0180$ , $p = 3.0 \times 10^{-37}$ |
| Number of cis-SNPs | 68.3%, $\Delta = +0.0174$ , $p = 1.8 \times 10^{-38}$ | 70.7%, $\Delta = +0.0197$ , $p = 2.0 \times 10^{-43}$ | 65.8%, $\Delta = +0.0196$ , $p = 2.0 \times 10^{-35}$ |
| Number of LD blocks | 69.1%, $\Delta = +0.0180$ , $p = 3.8 \times 10^{-40}$ | 69.8%, $\Delta = +0.0200$ , $p = 1.0 \times 10^{-43}$ | 65.7%, $\Delta = +0.0186$ , $p = 5.5 \times 10^{-34}$ |
| Gene-window size | 69.3%, $\Delta = +0.0196$ , $p = 2.2 \times 10^{-44}$ | 69.1%, $\Delta = +0.0190$ , $p = 1.0 \times 10^{-40}$ | 66.4%, $\Delta = +0.0181$ , $p = 4.2 \times 10^{-33}$ |

Supplementary Figure S2 plots the underlying per-gene $\Delta r$ distributions for the cis-SNP-count, block-count, and within-block-LD stratifications shown in Supplementary Table S4, confirming that the central tendency stays positive in each tertile rather than being driven by a tail of extreme genes.

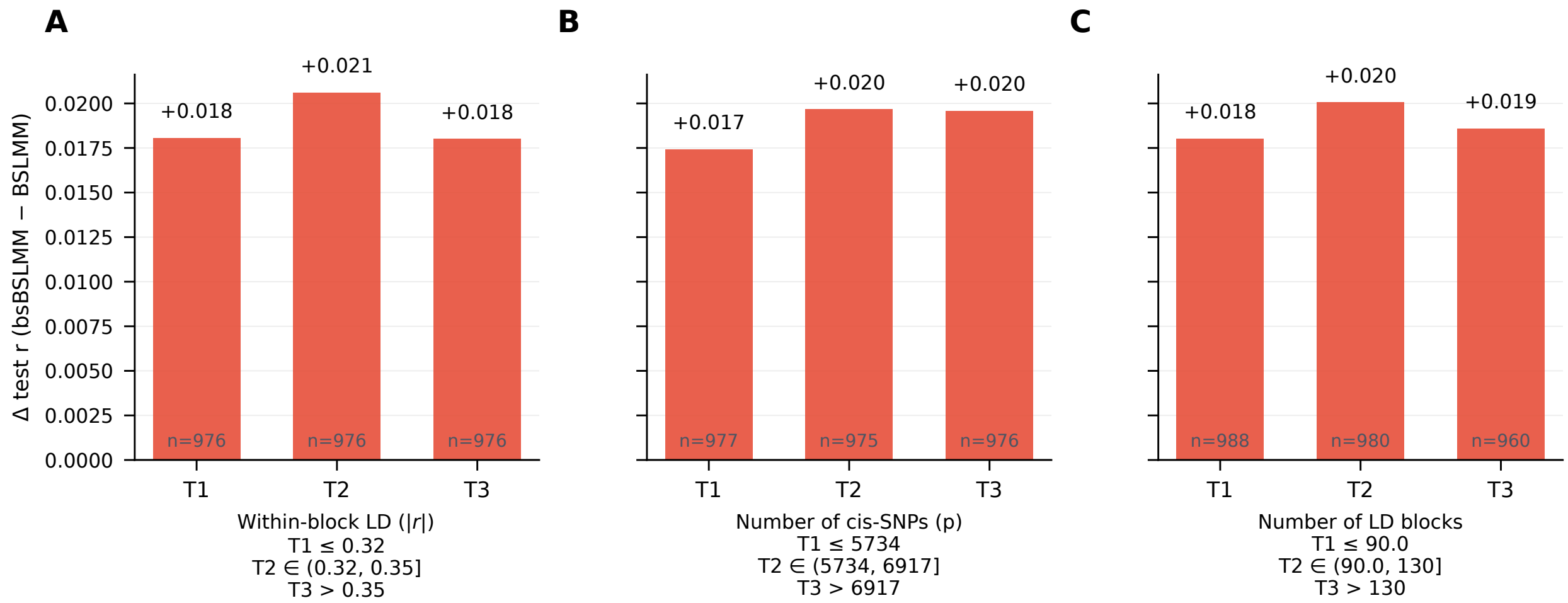

**Supplementary Figure S2.** Stratified held-out $\Delta r$ by SNP count, block count, and within-block LD. Source figure: manuscript/figures/Fig_S2_stratified_h2h.pdf.

## S3. Simulation analyses

To test the bsBSLMM prior under controlled data-generating mechanisms, we constructed three scenarios that span the architectures the model is expected to encounter in real cis-regulatory data. Supplementary Table S5 names each scenario, states what it is designed to probe, and gives the directional expectation that follows from the bsBSLMM hierarchy: a block-local sparse regime in which block gating should help, a non-local sparse regime in which the advantage should narrow, and a diffuse polygenic regime in which dense baselines should be competitive.

**Supplementary Table S5.** Simulation scenarios.

| Scenario | Purpose | Expected behavior |
|---|---|---|
| Sparse block-local effects | Tests whether block gating improves recovery when effects cluster in LD neighborhoods | bsBSLMM is expected to improve over BSLMM |
| Sparse non-local effects | Tests robustness when sparsity is present but block localization is weaker | bsBSLMM advantage is expected to narrow |
| Diffuse polygenic effects | Tests whether block sparsity over-penalizes diffuse architecture | BSLMM/BLUP-like behavior should be competitive |

Supplementary Table S6 reports the observed per-scenario means across 40 chr20 genes for 5-fold CV Pearson $r$ and coefficient-recovery correlation against the simulation truth. The pattern matches Supplementary Table S5's expectations: in the sparse-high-SNR regime bsBSLMM achieves the highest CV $r$ (0.605) and coefficient recovery (0.353), ahead of BSLMM (0.598 / 0.295) and LASSO (0.542 / 0.209); in the sparse-low-SNR regime the methods converge in prediction but coefficient recovery still favours bsBSLMM; in the polygenic-dominant regime BLUP wins on CV $r$ as expected and the bsBSLMM gain over BSLMM evaporates. Supplementary Figure S3 plots the same per-gene values that the means in Supplementary Table S6 summaries.

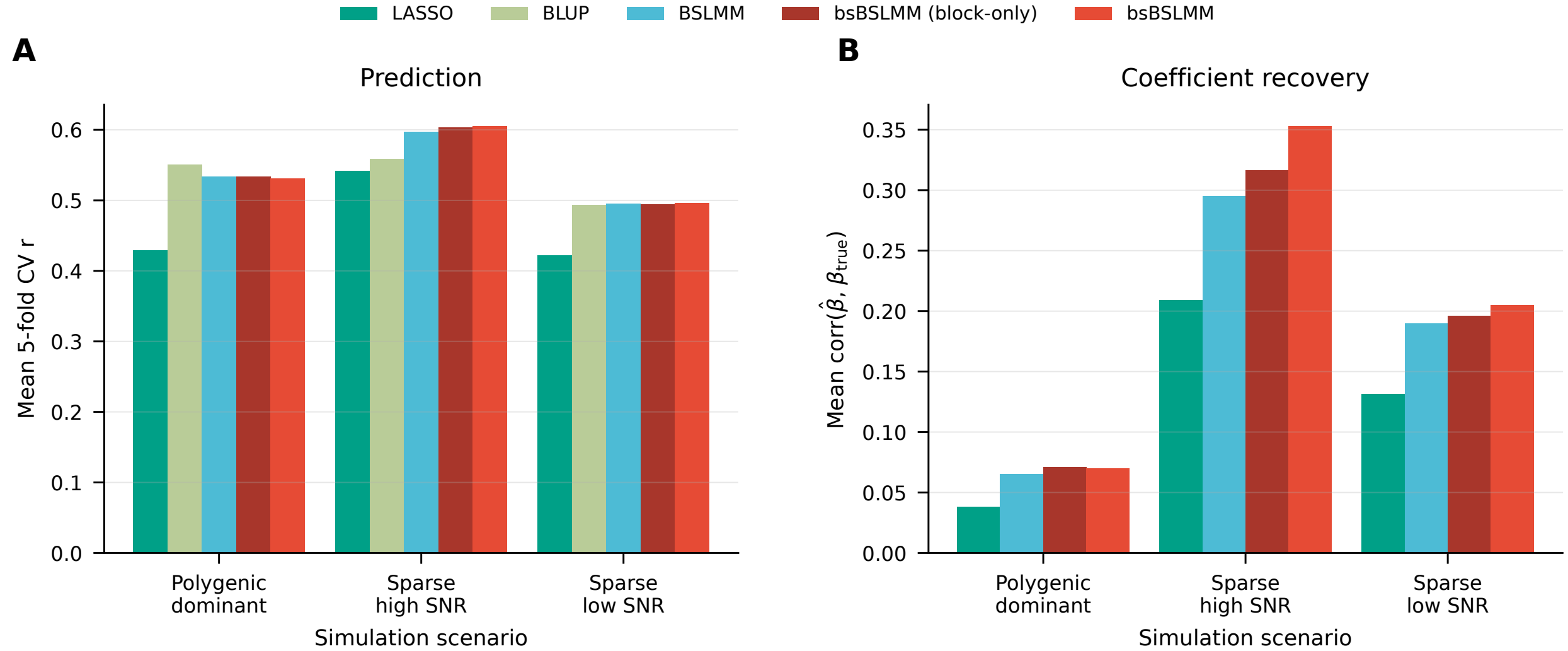

**Supplementary Figure S3.** Per-gene simulation recovery and prediction summaries across the three scenarios in Supplementary Table S5 (sparse high SNR, sparse low SNR, polygenic dominant), 40 chr20 genes per scenario. **S3a**: 5-fold CV Pearson $r$ per gene per method — the per-gene values whose per-scenario means appear in the "CV r" column of Supplementary Table S6. **S3b**: coefficient-recovery Pearson $r$ between the fitted sparse-effect vector and the simulation truth, per gene per method — the per-gene values whose per-scenario means appear in the "Coefficient recovery r" column of Supplementary Table S6. BLUP is shown in S3a but omitted from S3b because BLUP is a dense predictor and does not produce a sparse-effect estimate to compare against truth. Source figures: manuscript/figures/Fig_S3a_sim_cv_r.pdf (data: results/benchmark_G/analysis_simulation/sim_long.csv) and manuscript/figures/Fig_S3b_sim_beta_corr.pdf (same source).

**Supplementary Table S6.** Simulation prediction and coefficient-recovery summaries. Values are means across 40 chr20 genes per scenario. CV performance is 5-fold CV Pearson $r$. Coefficient recovery is the Pearson correlation between the fitted sparse-effect vector and the simulation truth; it is not reported for BLUP because BLUP is a dense predictor rather than a sparse-effect estimator. Source data: results/benchmark_G/analysis_simulation/sim_long.csv.

| Scenario | Method | Genes | Mean CV Pearson $r$ | Mean coefficient recovery |
|---|---|---|---|---|
| Sparse high SNR | LASSO | 40 | 0.542 | 0.209 |
| Sparse high SNR | BLUP | 40 | 0.559 | - |
| Sparse high SNR | BSLMM | 40 | 0.598 | 0.295 |
| Sparse high SNR | bsBSLMM block-only | 40 | 0.603 | 0.317 |
| Sparse high SNR | bsBSLMM | 40 | 0.605 | 0.353 |
| Sparse low SNR | LASSO | 40 | 0.422 | 0.132 |
| Sparse low SNR | BLUP | 40 | 0.494 | - |
| Sparse low SNR | BSLMM | 40 | 0.495 | 0.190 |
| Sparse low SNR | bsBSLMM block-only | 40 | 0.495 | 0.196 |
| Sparse low SNR | bsBSLMM | 40 | 0.497 | 0.205 |
| Polygenic dominant | LASSO | 40 | 0.429 | 0.038 |
| Polygenic dominant | BLUP | 40 | 0.551 | - |
| Polygenic dominant | BSLMM | 40 | 0.534 | 0.065 |
| Polygenic dominant | bsBSLMM block-only | 40 | 0.534 | 0.071 |
| Polygenic dominant | bsBSLMM | 40 | 0.531 | 0.070 |

## S4. Ablation details and posterior interpretation

Table 2 of the main text reports the headline +388-gene yield gain of full bsBSLMM over BSLMM, decomposed across the block layer and the TSS prior. Supplementary Table S7 expands that summary into the complete three-row strata used to compute the decomposition, listing the absolute counts of genes passing both prediction filters for BSLMM, block-only bsBSLMM, and full bsBSLMM, together with the shared-gene counts and paired Wilcoxon means for each pairwise comparison.

**Supplementary Table S7.** Genome-wide ablation strata under both prediction filters.

| Model | Genes passing both filters | Comparison set | Shared genes | Higher $r$ for newer model | Mean $\Delta r$ |
|---|---|---|---|---|---|
| BSLMM | 3,130 | - | - | - | - |
| bsBSLMM block-only | 3,395 | block-only vs BSLMM | 2,922 | 1,914 | +0.0155 |
| bsBSLMM | 3,518 | bsBSLMM vs block-only | 3,248 | 1,861 | +0.0035 |
| bsBSLMM | 3,518 | bsBSLMM vs BSLMM | 2,928 | 1,998 | +0.0189 |

Reading down the genes-passing-both-filters column of Supplementary Table S7, the 388-gene yield increase decomposes into +265 genes from the block layer (BSLMM → block-only) and +123 genes from the TSS-distance prior (block-only → full). The corresponding paired-mean improvements, +0.0155 and +0.0035, sum to +0.0190, matching the direct bsBSLMM-vs-BSLMM estimate of +0.0189 up to rounding — so the two components contribute additively in this dataset. Supplementary Figure S4 plots the same decomposition as a bar chart so the yield gain and the per-stratum paired held-out summary can be read off briefly; this is the extended visualization referenced from §3.3 of the main text.

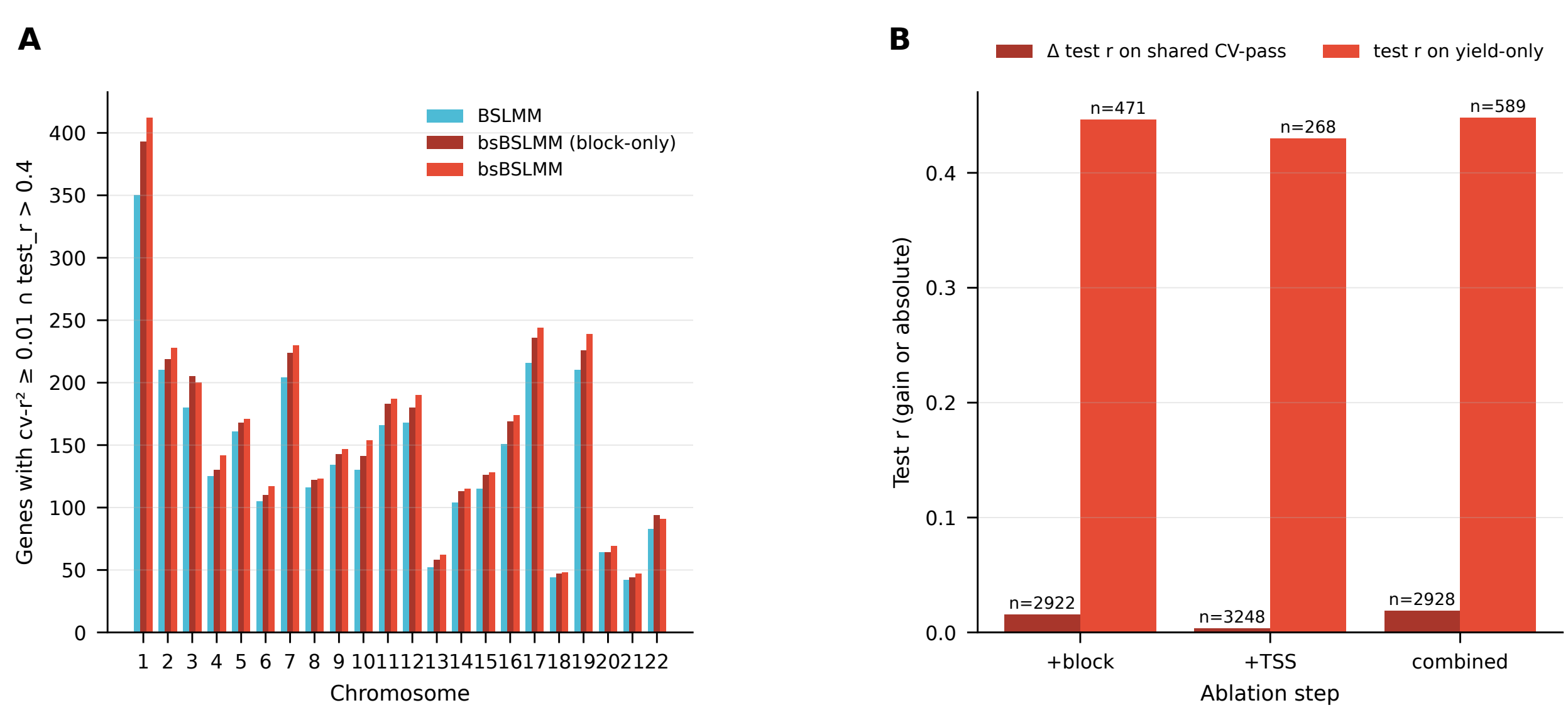

**Supplementary Figure S4.** Ablation of LD-block sparsity and the TSS-distance prior. The bars decompose the 388-gene bsBSLMM gain over BSLMM into the +265 genes contributed by the block layer (BSLMM → block-only) and the +123 genes contributed by the TSS-distance prior (block-only → full), with the matching paired held-out improvements shown alongside. Source figure: manuscript/figures/Fig_S4_ablation.pdf.

The main-text posterior-architecture and TSS-prior diagnostics (main-text Figure 3) are computed from results/benchmark_G/analysis_biology_stage1/arch_per_gene.csv: 5,714 of 23,098 genes (24.7%) had at least one credibly active SNP; among those genes, the median sparse genetic-variance share was PGE = 0.62; and the learned TSS-distance coefficient was negative in 22,812 of 23,098 genes (98.8%). To make the locus-level shape of this posterior concrete, Supplementary Figure S5 compares bsBSLMM and BSLMM side by side on two case studies — NAPRT (chr8, a multi-SNP sparse-cis architecture in which the bsBSLMM block prior cleanly localises six credibly-active variants while BSLMM diffuses the signal across more SNPs at lower posterior support) and ERAP2 (chr5, the canonical single-effect cis-eQTL where both methods identify the same dominant variant). Each gene gets a stacked pair of panels: top = bsBSLMM posterior mean $\beta$ with 95% credible intervals, bottom = BSLMM posterior mean $\beta$ with $\gamma \geq 0.5$ highlighting. All cis-SNPs appear as faint grey dots; signed stems mark the actively included variants.

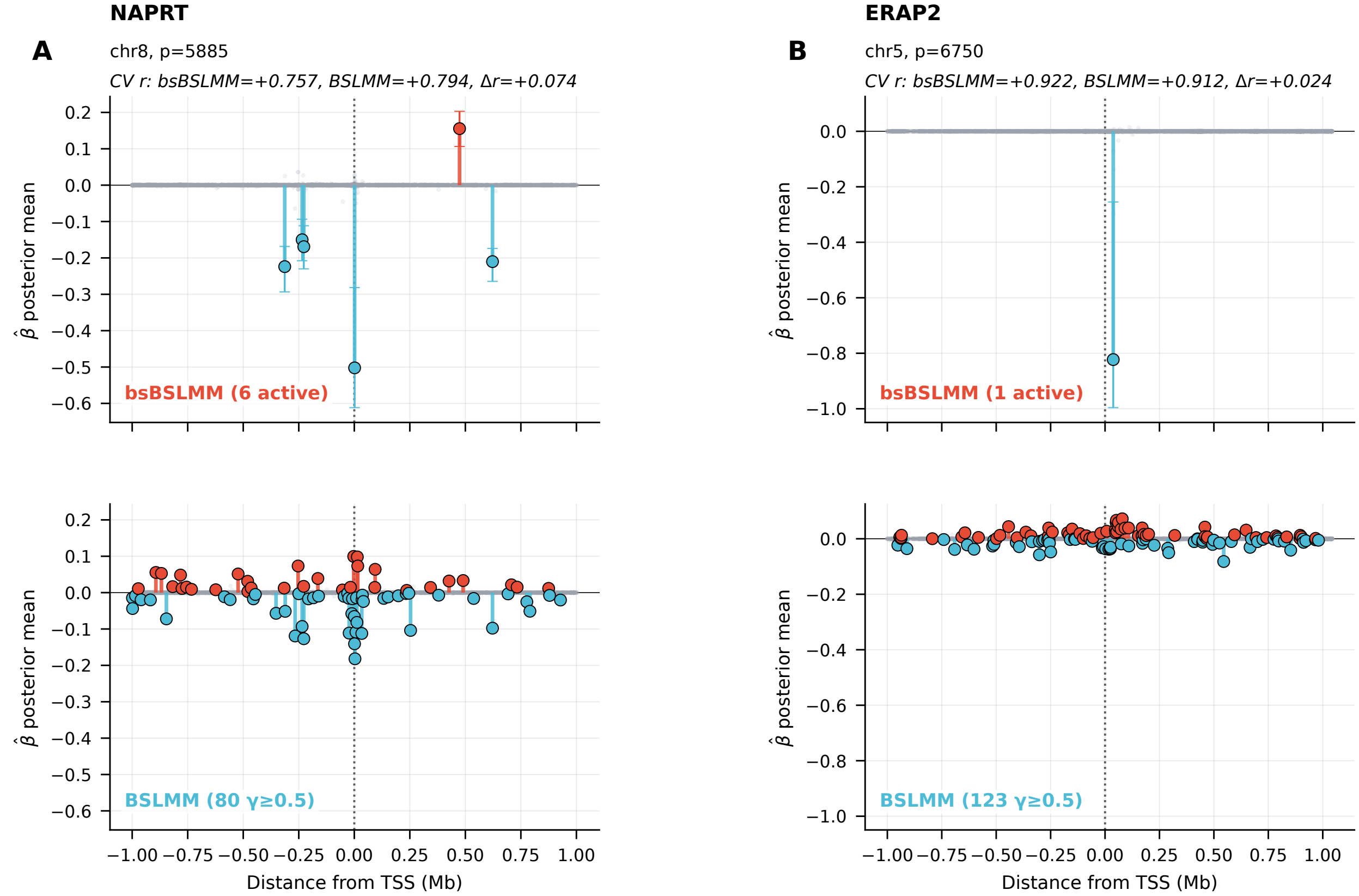

**Supplementary Figure S5.** Locus-level case studies comparing bsBSLMM (top row) against BSLMM (bottom row) for NAPRT (chr8, **A**) and ERAP2 (chr5, **B**). x-axis: SNP-to-TSS distance (Mb). y-axis: posterior mean β. Faint grey dots are all cis-SNPs; coloured stems with error bars are credibly-active SNPs in bsBSLMM (95% posterior interval excludes zero) and $\gamma \geq 0.5$ SNPs in BSLMM. Bar colour encodes sign of β (red = positive, cyan = negative). Source figure: manuscript/figures/Fig_S5_case_studies.pdf.

# S5. Functional enrichment and TWAS details

## S5.1 Active-set definitions

Active SNP sets were defined separately for each method to respect method-specific outputs. For Bayesian models, active status was based on posterior summaries of sparse effects. For penalized or external imputation baselines, active status followed non-zero or method-reported weight definitions available in the corresponding output files. The same cis-window universe was used when estimating fold enrichment.

## S5.2 Annotation enrichment

The main manuscript (Figure 4) reports the ENCODE regulatory tracks [9] and GTEx v8 LCL expression quantitative trait locus (eQTL) track [8]. The regulatory tracks are GM12878 DNase I hypersensitivity (DNase) peaks and histone H3 lysine 27 acetylation (H3K27ac) peaks. The genome-wide association study (GWAS) Catalog track [10] is retained here as a fourth, broader annotation. Supplementary Table S8 gives the full fold-enrichment estimate and bootstrap 95% confidence interval for each method × track combination, so all four tracks can be compared in one place.

**Supplementary Table S8.** Fold-enrichment of method-active cis-SNPs across all annotation tracks.

| Method | GWAS Catalog | GM12878 DNase | GM12878 H3K27ac | GTEx v8 LCL cis-eQTL |
|---|---|---|---|---|
| BSLMM | 1.17 [1.11, 1.23] | 1.40 [1.32, 1.50] | 1.35 [1.28, 1.42] | 2.58 [2.45, 2.72] |
| bsBSLMM | 1.33 [1.14, 1.55] | 3.04 [2.63, 3.49] | 2.02 [1.77, 2.29] | 2.78 [2.40, 3.18] |
| TIGAR EN | 2.18 [2.13, 2.22] | 1.52 [1.48, 1.56] | 1.46 [1.42, 1.50] | 4.21 [4.02, 4.39] |
| TIGAR DPR | 2.42 [2.37, 2.48] | 1.97 [1.90, 2.04] | 1.76 [1.70, 1.82] | 2.09 [2.00, 2.18] |

Two observations on the GWAS Catalog column of Supplementary Table S8 are worth flagging. First, the count is version-sensitive because the Catalog is updated over time, so the main text avoids treating GWAS Catalog enrichment as a central result. Second, TIGAR-style methods produce visibly larger active sets and consequently larger overlap with the Catalog regardless of whether the Catalog overlap is regulatory or merely incidental, which is why the LCL-specific DNase and H3K27ac tracks are the more informative comparison.

### S5.3 IBD TWAS

The inflammatory bowel disease (IBD) transcriptome-wide association study (TWAS) used de Lange et al. summary statistics [11] and assessed both calibration and immune-locus recovery.

Selected bsBSLMM IBD signals included IL23R, ZPBP2, PTPN2 as reported by de Lange et al. [11] and Liu et al. [12], and ATG16L1 as reported by Hampe et al. [13] and Rioux et al. [14], all of which are consistent with established IBD biology. bsBSLMM contributed 30 method-unique genome-wide significant ($p < 5 \times 10^{-8}$) genes across 12 chromosomes. These 30 genes aggregated 1,717-5,101 allele-matched GWAS SNPs per gene through the posterior-mean weight vector. Source files: results/benchmark_G/analysis_biology_stage3/twas_summary.csv and results/benchmark_G/analysis_biology_stage3/top_bsbslmm_unique_hits.csv.

### S5.4 LOS prediction benchmark

The LOS prediction benchmark used cohort-specific matched-power prediction filters, requiring CV-$r^2 \geq 0.01$ and held-out test $r > 0.35$ in CA or $r > 0.43$ in AA. These thresholds were chosen so the two-sided correlation test at the cohort median held-out sample size approximated the GEUVADIS reference threshold. Source report: results/analysis_panel_los_v02/los_h2h_matched_power.md.

Supplementary Table S9 reports the per-method count passing both LOS prediction filters together with the CV-pass denominator and the retention ratio, separately for the CA and AA cohorts. The headline LOS counts cited in §3.6 (CA 1,080; AA 722) come from this table, and the per-method ordering matches what was observed on GEUVADIS: bsBSLMM has the largest pass-both-filters count in both cohorts, followed by TIGAR DPR, BSLMM, LASSO, TIGAR EN, and BLUP.

**Supplementary Table S9.** LOS matched-power prediction yield.

| Cohort | Method | Genes passing both filters | CV-pass genes | Retention |
|---|---|---|---|---|
| CA | bsBSLMM | 1,080 | 15,478 | 7.0% |
| CA | BSLMM | 955 | 16,234 | 5.9% |
| CA | LASSO | 903 | 6,150 | 14.7% |
| CA | BLUP | 548 | 8,186 | 6.7% |
| CA | TIGAR EN | 852 | 6,814 | 12.5% |
| CA | TIGAR DPR | 959 | 19,477 | 4.9% |
| AA | bsBSLMM | 722 | 19,472 | 3.7% |
| AA | BSLMM | 659 | 23,720 | 2.8% |
| AA | LASSO | 476 | 8,934 | 5.3% |
| AA | BLUP | 254 | 11,551 | 2.2% |
| AA | TIGAR EN | 399 | 9,728 | 4.1% |
| AA | TIGAR DPR | 604 | 25,775 | 2.3% |

Yield, however, only counts gene-level coverage. To compare per-gene prediction accuracy among methods that share a gene, Supplementary Table S10 reports the paired held-out Wilcoxon comparison of bsBSLMM against each of the five baselines, separately for CA and AA. In CA the mean $\Delta r$ favours bsBSLMM against every baseline ($+0.003$ to $+0.044$, four of five with $p < 0.01$); in AA the direction is the same but only BSLMM, LASSO, and BLUP reach $p < 0.05$, consistent with the reduced sample size of the AA cohort.

**Supplementary Table S10.** LOS paired held-out comparison under the matched-power prediction filters.

| Cohort | Comparison | Shared genes | Higher $r$ for bsBSLMM | Mean $\Delta r$ | Wilcoxon $p$ |
|---|---|---|---|---|---|
| CA | bsBSLMM vs BSLMM | 703 | 468 (66.6%) | +0.0188 | $1.22 \times 10^{-24}$ |
| CA | bsBSLMM vs LASSO | 765 | 403 (52.7%) | +0.0027 | 0.0477 |
| CA | bsBSLMM vs BLUP | 405 | 284 (70.1%) | +0.0442 | $6.35 \times 10^{-25}$ |
| CA | bsBSLMM vs TIGAR EN | 674 | 371 (55.0%) | +0.0060 | 0.00111 |
| CA | bsBSLMM vs TIGAR DPR | 653 | 387 (59.3%) | +0.0115 | $7.25 \times 10^{-9}$ |
| AA | bsBSLMM vs BSLMM | 364 | 203 (55.8%) | +0.0088 | 0.000803 |

| Cohort | Comparison | Shared genes | Higher $r$ for bsBSLMM | Mean $\Delta r$ | Wilcoxon $p$ |
|---|---|---|---|---|---|
| AA | bsBSLMM vs LASSO | 333 | 188 (56.5%) | +0.0040 | 0.0227 |
| AA | bsBSLMM vs BLUP | 120 | 72 (60.0%) | +0.0298 | 0.00118 |
| AA | bsBSLMM vs TIGAR EN | 260 | 148 (56.9%) | +0.0042 | 0.0699 |
| AA | bsBSLMM vs TIGAR DPR | 281 | 147 (52.3%) | +0.0057 | 0.171 |

## S5.5 LOS TWAS and GSEA

After downstream S-PrediXcan-style TWAS against the Morris et al. estimated bone mineral density (eBMD) GWAS [15], all four cohort × method combinations show appreciable pre-calibration inflation (1.90-2.12), so we apply post-hoc genomic control [16] and report calibrated hit counts. Supplementary Table S11 gives the number of genes tested, the post-GC Bonferroni-significant hit count, and the pre-GC $\lambda_{GC}$ per cohort and method. CA bsBSLMM (151 hits) exceeds CA BSLMM (116) by 35 genes, while AA bsBSLMM and AA BSLMM produce comparable hit counts (177 vs 181) but bsBSLMM enters the GC step with lower inflation (1.90 vs 2.12).

**Supplementary Table S11.** LOS S-PrediXcan Bonferroni-significant hits against Morris estimated bone mineral density (eBMD) GWAS after post-hoc genomic control [16].

| Cohort | Method | Genes tested | Bonferroni hits after GC | Pre-GC $\lambda_{GC}$ |
|---|---|---|---|---|
| CA | BSLMM | 16,191 | 116 | 1.98 |
| CA | bsBSLMM | 15,416 | 151 | 1.95 |
| AA | BSLMM | 23,636 | 181 | 2.12 |
| AA | bsBSLMM | 19,395 | 177 | 1.90 |

To probe biological coherence rather than the per-gene significance count alone, Supplementary Table S12 reports preranked GSEA [17] against three curated BMD-relevant gene sets (Morris-BMD [15], BMD-TWAS-coloc [18], and LOS-SV-TWAS [19]) for each cohort × method combination, showing the top-scoring gene set, its normalised enrichment score, and FDR $q$. Three of the four combinations recover the same Morris-derived BMD-TWAS-colocalization set; the most striking signal — AA bsBSLMM on the LOS structural-variant TWAS set ($\text{NES} = -1.89$, FDR $q = 0.001$) — does not have an AA-BSLMM counterpart that passes the same FDR threshold, providing the population-matched evidence highlighted in the main text.

**Supplementary Table S12.** Curated BMD preranked-GSEA results.

| Cohort | Method | Gene set | Normalized enrichment score (NES) | False-discovery-rate (FDR) $q$ |
|---|---|---|---|---|
| CA | BSLMM | G_BMD-TWAS-coloc | -1.70 | 0.018 |
| CA | bsBSLMM | G_BMD-TWAS-coloc | -1.72 | 0.022 |
| AA | BSLMM | G_BMD-TWAS-coloc | -1.45 | 0.036 |
| AA | bsBSLMM | G_LOS-SV-TWAS | -1.89 | 0.001 |

## S7. Runtime and computational feasibility

The runtime experiment quantifies computational feasibility under the matched fitting protocol. Supplementary Table S13 reports, for each method, the median per-gene full-fit wall time on chr20, the corresponding throughput at $J = 60$ concurrent slots, and the genome-wide mean held-out test $r$ among CV-pass genes — three numbers chosen to expose both the absolute cost of each method and the quality it buys.

**Supplementary Table S13.** Per-gene full-fit wall time and throughput at $J = 60$ on chr20.

| Method | Median wall time per gene (s) | Approximate throughput at $J = 60$ (genes/hour) | Genome-wide mean test $r$ among CV-pass genes |
|---|---|---|---|
| BLUP | 0.0115 | not limiting | 0.5447 |
| LASSO | 13.97 | 15,463 | 0.5749 |
| BSLMM | 34.44 | 6,271 | 0.5677 |
| bsBSLMM (Numba) | 55.59 | 3,885 | 0.5706 |
| bsBSLMM (C++) | 30.81 | 7,011 | 0.5706 |

The MCMC methods in Supplementary Table S13 used matched schedules ($n_{\text{iter}} = 5{,}000$, $n_{\text{burn}} = 2{,}500$, $n_{\text{thin}} = 25$); the chr20 benchmark used 10 genes by 3 seeds per method under single-threaded BLAS. The native C++ block-sweep variant is reported separately from the Python+Numba path because it removes the wall-time premium relative to matched-MCMC BSLMM (30.8 s vs 34.4 s per gene) while preserving the same statistical model and the same held-out test $r$. LASSO and BLUP remain substantially faster than either MCMC variant. Supplementary Figure S7 extends Supplementary Table S13 along two further axes: panel A pairs the chr20 wall-time medians with the genome-wide Table 1 count of genes passing both prediction filters to form a cost-yield frontier across methods, and panel B uses a separate 20-gene benchmark drawn log-uniformly across the genome-wide cis-window range (74 to 30,012 cis-SNPs) to estimate empirical wall-time scaling exponents, providing a sense of how the per-gene cost in Supplementary Table S13 generalizes to much wider windows.

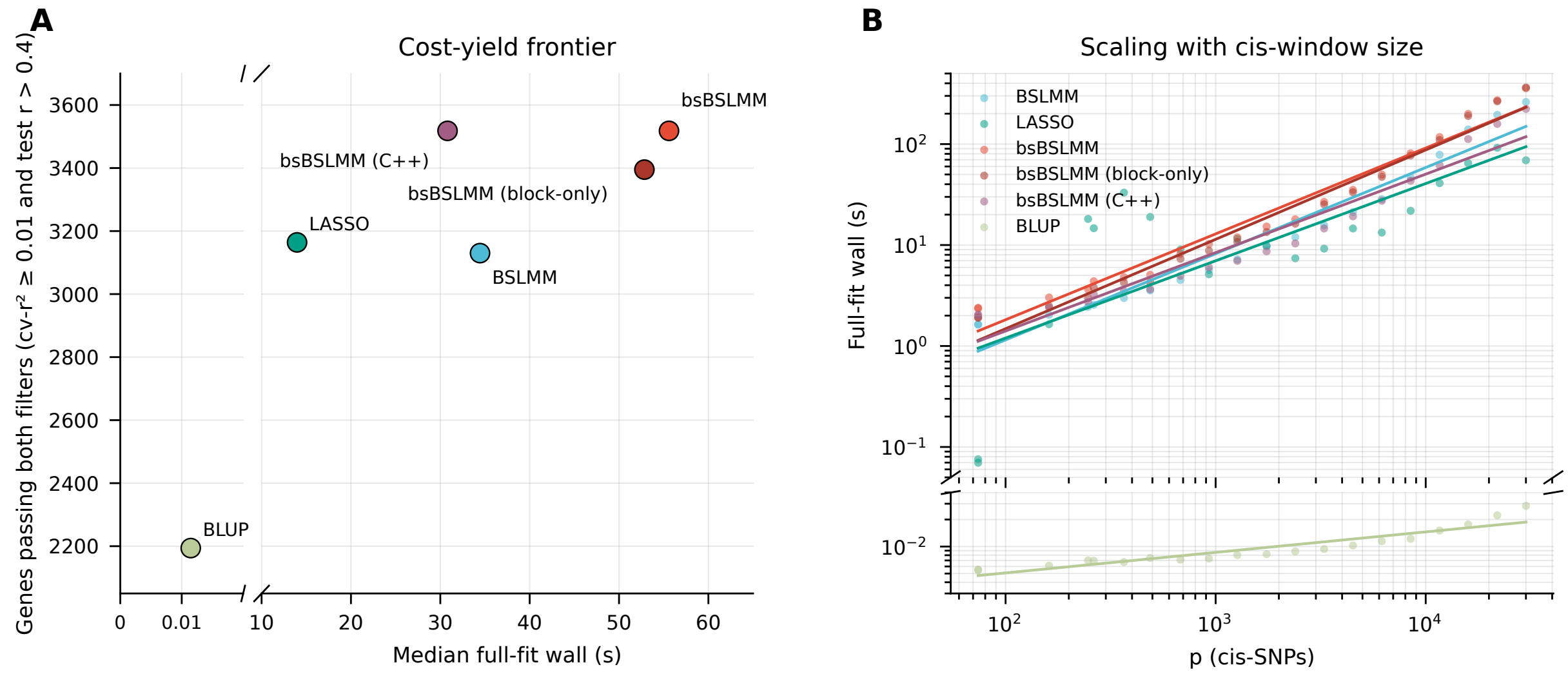

**Supplementary Figure S7.** Runtime per method. **A**: cost-yield frontier — median full-fit wall time on chr20 (x-axis) versus the genome-wide count of genes passing both prediction filters (y-axis, the Table 1 yield: CV-$r^2$ $\geq$ 0.01 and held-out r > 0.4). This count is used in place of mean held-out test r among CV-pass genes because each method's CV-pass set has a different size (LASSO 8,387; bsBSLMM 12,236; Table 1, "CV-only pass" column), and averaging test r over each method's own smaller subset induces a survivor-bias artefact that inflates the apparent quality of the more aggressive screeners. The count is sample-size-free and matches the headline retention claim in §3.1. The x-axis is a broken linear scale dedicated to the two natural wall-time clusters (BLUP alone near 11 ms, dominated by the GRM build and eigendecomposition; LASSO, BSLMM, bsBSLMM (block-only), bsBSLMM (C++), and bsBSLMM (Numba) in the 14–56 s range). The bsBSLMM (C++) variant is shown in NPG muted purple so it remains distinguishable from the bsBSLMM Numba (red) and BSLMM (cyan) points that sit at nearby coordinates. TIGAR EN/DPR are not shown because the chr20 timing CSV does not include matched wall times for the TIGAR pipeline. **B**: wall-time scaling with cis-window size on broken log-log axes, with per-method power-law fits. The scaling panel uses a separate run on 20 protein-coding genes sampled log-uniformly across the genome-wide *p* range (74 to 30,012 cis-SNPs, ~2.6 $\log_{10}$-decades), three seeds per gene, so the fitted slopes are interpretable as empirical scaling exponents rather than noise lines over the narrow *p* band (~0.3 $\log_{10}$-decades) of the chr20 cost-yield sample. BLUP is shown in a separate lower band because its wall time sits two to four orders of magnitude below the MCMC and LASSO cluster. Source figure: manuscript/figures/Fig_S7_runtime.pdf (data: results/benchmark_G/analysis_speed/speed_chr20.csv, results/benchmark_G/analysis_panel/panel_long.csv, and results/benchmark_G/analysis_speed/speed_scaling.csv).

## S8. Source data and computational reproducibility

Supplementary Table S14 lists the aggregated source files and scripts used to generate the manuscript figures, tables, and extended supplementary results. Paths are relative to the project repository.

**Supplementary Table S14.** Source data and scripts for the main result families.

| Result family | Main-text location | Extended supplementary location | Source data | Analysis script |
|---|---|---|---|---|
| GEUVADIS yield and paired benchmark | Table 1; Figure 2 | Tables S2-S4 | results/benchmark_G/analysis_panel_t40/ | scripts/panel_compare.py |
| Ablation | Table 2 | Table S7; Figure S4 | results/benchmark_G/analysis_ablation_t40/ | scripts/ablation_compare.py |
| Simulation | §3.1 | Tables S5-S6; Figure S3 | results/benchmark_G/analysis_simulation/sim_long.csv | scripts/simulation.py; scripts/merge_sim_chunks.py; scripts/make_figures.py |
| Posterior architecture and TSS prior | Figure 3; §3.4 | Figure S5 | results/benchmark_G/analysis_biology_stage1/; manuscript/figures/Fig_03_arch_tss.pdf | scripts/biological_analysis_stage1.py |
| Functional enrichment | Figure 4; Table 3 | Table S8 | results/benchmark_G/analysis_biology_stage2/ | scripts/biological_analysis_stage2.py |
| IBD TWAS | Table 3 | §S5.3 | results/benchmark_G/analysis_biology_stage3/ | scripts/biological_analysis_stage3.py |
| LOS prediction benchmark | §3.6 | Tables S9-S10 | results/analysis_panel_los_v02/ | los/scripts/05b_panel_compare_v02.py; los/scripts/05c_recompute_h2h_matched_power.py |
| LOS TWAS and GSEA | Table 3 | Tables S11-S12 | results/spredixcan_los_gc/; results/gsea_los/ | los/scripts/10b_genomic_control.py; los/scripts/12_gsea_preranked.py |
| Runtime | §3.7 | Table S13; Figure S7 | results/benchmark_G/analysis_speed/speed_chr20.csv; results/benchmark_G/analysis_panel/panel_long.csv; results/benchmark_G/analysis_speed/speed_scaling.csv (20 genes, p ∈ [74, 30,012], 3 seeds) | scripts/speed_benchmark.py; scripts/make_figures.py |

The repository also contains PIPELINE.md, which provides command-level replication instructions for rebuilding the benchmark panels, downstream analyses, and figures from the source data.

## Supplementary References